\documentclass[english,superscriptaddress,onecolumn,aps,prc]{revtex4-2}
\usepackage{graphicx} % Required for inserting images
\usepackage{amsmath}
\usepackage{slashed}
\usepackage{comment}
\usepackage{xcolor}

\newcommand{\ii}{i}

\def\wT{{\widehat T}}
\def\wj{{\widehat j}}

\def\wP{{\widehat P}}

\def\wQ{{\widehat Q}}

\def\wrhol{{\widehat{\rho}_{\rm LE}}}

\def\wa{\widehat a}
\def\wad{\widehat a^{\dagger}}

\newcommand{\tr}{{\rm tr}}  
\newcommand{\Tr}{{\rm Tr}}

\newcommand{\e}{{\rm e}}
\newcommand{\p}{{\rm p}}

\newcommand{\y}{{\rm y}}

\newcommand{\di}{{\rm d}}
\newcommand{\be}{\begin{equation}}
\newcommand{\ee}{\end{equation}}                                                                               
\def\bea{\begin{eqnarray}}
\def\eea{\end{eqnarray}} 
\usepackage[normalem]{ulem}
\usepackage[thicklines]{cancel}

\begin{document}

\title{An improved formula for Wigner function and spin polarization in 
a decoupling relativistic fluid at local thermodynamic equilibrium}

\author{Xin-Li Sheng}
\email{sheng@fi.infn.it}
\affiliation{Universit\`a degli studi di Firenze and INFN Sezione di Firenze,\\
Via G. Sansone 1, I-50019 Sesto Fiorentino (Florence), Italy}

\author{Francesco Becattini}
\email{becattini@fi.infn.it}
\affiliation{Universit\`a degli studi di Firenze and INFN Sezione di Firenze,\\
Via G. Sansone 1, I-50019 Sesto Fiorentino (Florence), Italy}

\author{Daniele Roselli}
\email{daniele.roselli@unifi.it}
\affiliation{Universit\`a degli studi di Firenze and INFN Sezione di Firenze,\\
Via G. Sansone 1, I-50019 Sesto Fiorentino (Florence), Italy}

\begin{abstract}
We present an upgraded formula for Wigner function and spin polarization of fermions emitted 
by a relativistic fluid at local thermodynamic equilibrium at the decoupling which improves 
the one obtained in refs. \cite{Becattini:2021suc,Liu:2021uhn} and used in numerical simulations 
of relativistic nuclear collisions. By using a new expansion method, applicable to decoupling 
hypersurfaces with arbitrary geometry, we reproduce the known term proportional to thermal vorticity 
and obtain an upgraded form of the spin-shear term which captures the dependence on the geometry. 
The new method also includes additional contributions whose physical nature is to be assessed. 
The new expression also naturally excludes contributions from space-time gradients in the normal 
direction of the hypersurface, providing a theoretical justification for the isothermal condition 
previously imposed {\em a priori}. This framework can be extended to particles with arbitrary spin.
\end{abstract}

\maketitle

%********************************************************************************
\section{Introduction.}
%********************************************************************************

Relativistic heavy-ion collisions provides a unique environment for studying strongly interacting 
matter, known as the quark-gluon plasma (QGP), under extremely high temperature \citep{Collins:1974ky,Shuryak:1977ut,Busza:2018rrf}. 
After a short pre-equilibrium evolution, the system is assumed to achieve local thermodynamic 
equilibrium (LTE) and its evolution is thereafter effectively described by relativistic hydrodynamics \citep{Stoecker:1986ci,Rischke:1995ir,Rischke:1995mt,Jaiswal:2016hex,Jaiswal:2020hvk}. 
Physical quantities are then characterized by slowly varying macroscopic variables, such as flow velocity, 
temperature, and chemical potential. The non-equilibrium operator's approach and its reformulation \citep{Zubarev:1966,Zubarev:1979afm,Zubarev:1989su,VANWEERT1982133,Becattini:2019dxo} offers a rigorous 
framework to obtain physical observables in a relativistic fluid based on quantum statistical field theory 
at local equilibrium, which is especially useful for spin \citep{Becattini:2013fla,Becattini:2019dxo,Becattini:2020sww,Becattini:2021iol,Becattini:2021suc,Liu:2021nyg,Sheng:2024pbw,Zhang:2024mhs,Li:2025pef,Buzzegoli:2025zud}, an inherently quantum observable  \citep{STAR:2017ckg,STAR:2018gyt,STAR:2019erd,ALICE:2019onw,STAR:2022fan,Liang:2004ph,Becattini:2022zvf,Becattini:2024uha,Liang:2004xn,Chen:2023hnb,Chen:2024afy}.

If the QGP achieves LTE on some initial hypersurface $\Sigma_0$, the density operator is obtained by maximizing the entropy 
$S=-\text{tr}(\widehat\rho\,\text{log}\widehat\rho)$ with constrained energy-momentum and charge densities. With the 
Belinfante stress-energy operator $\widehat{T}^{\mu\nu}$ \cite{BELINFANTE1940449} and the current operator $\widehat{j}^\mu$, 
the density operator takes the following form \cite{Zubarev:1966,VANWEERT1982133,Becattini:2019dxo}:
\begin{equation}\label{LE-density-operator}
\widehat{\rho}=\frac{1}{Z}\exp\left[-\int_{\Sigma_0} \!\!\!\di\Sigma_{\mu}(y)\left(\widehat{T}^{\mu\nu}(y)\beta_{\nu}(y)\!-\!{\wj^\mu(y)\zeta(y)}\right)\right]\;,
\end{equation}
where $\beta$ is the four-temperature vector and $\zeta$ the reduced chemical potential $\mu/T$. By means of 
the Gauss' theorem, the exponent in the operator \eqref{LE-density-operator} can be transformed into the sum of a 3D integral 
at the decoupling hypersurface $\Sigma_{\rm D}$, that is when the fluid breaks up,
\footnote{In this work the stage at which the fluid ceases to exist is called decoupling. Freeze-out is the stage when 
interactions between hadrons cease and it is conceptually different from the former.}
and a 4D integral over the region encompassed by $\Sigma_{\rm D}$ and $\Sigma_0$, which is essentially the region of existence 
of the QGP \cite{Becattini:2019dxo}:
\begin{equation}\label{gauss}
\int_{\Sigma_{\rm D}} \!\! \di \Sigma_{\mu}(y) \; \left(\widehat{T}^{\mu\nu}(y)\beta_{\nu}(y) -\wj^\mu(y)\zeta(y)\right)+ 
\int_\Omega \di^4 y \; \left(\widehat{T}^{\mu\nu}(y) \partial_\mu \beta_\nu(y)-\wj^\mu(y)\partial_\mu\zeta(y)\right)\;.
\end{equation}
While the first term is the LTE integral at the decoupling, the second term accounts for the dissipative terms of physical 
observables. In this work, we will confine ourselves to the calculation of observables, notably spin polarization, at LTE at the decoupling. The dissipative terms are assumed to be small compared to the LTE part, so that we will neglect the second integral 
in the eq. \eqref{gauss} and focus on the zeroth order approximation, i.e. the LTE density operator:
\begin{equation}\label{LE-density-operator-new}
\widehat{\rho}_{\rm LE}=\frac{1}{Z}\exp\left[-\int_{\Sigma_{\rm D}} \!\!\!\di\Sigma_{\mu}(y)
\left(\widehat{T}^{\mu\nu}(y)\beta_{\nu}(y)-\wj^\mu(y)\zeta(y)\right)\right]\,.
\end{equation}

The mean value of a local operator $\widehat{O}(x)$ on a point $x \in \Sigma_{\rm D}$ calculated with the density operator 
\eqref{LE-density-operator-new}, e.g.
\be\label{LTEO}
 O_{\rm LE}(x) \equiv \Tr (\widehat{O}(x) \, \wrhol)\;,
\ee
should meet a definite requirement: it cannot depend on the values of the $\beta$ and $\zeta$ fields outside the
hypersurface $\Sigma_{\rm D}$. The reason is clear: the density operator involves an integration over $\Sigma_{\rm D}$ and 
since $x$ itself is a point on $\Sigma_{\rm D}$, the functional form of these fields outside this hypersurface can 
be varied arbitrarily without affecting the result. This requirement has a remarkable implication: in a gradient
expansion of the mean value at LTE, only the components along the tangent directions to $\Sigma_{\rm D}$ matter, the 
component perpendicular to the hypersurface should not appear in the final result. 

In previous spin polarization calculations with \eqref{LE-density-operator-new}, this requirement was not 
met, due to significant geometric approximations, tacitly or explicitely introduced \cite{Becattini:2021suc, Liu:2021uhn}. 
For instance, in ref. \cite{Becattini:2021suc} for the purpose of calculating the LTE contribution, the region encompassed
by the initial and the decoupling hypersurface was approximated with a rectangular region with infinitely distant
time-like boundaries, while in ref. \cite{Liu:2021uhn} the LTE hypersurface was tacitly assumed to be perpendicular to the 
four-velocity field, which is not the case in general. In fact, according to analytical models 
\cite{Bjorken:1982qr,Luzum:2008cw,Romatschke:2017ejr} and numerical simulations \cite{Song:2010aq,Shen:2014vra,Plumberg:2015eia,Elfner:2022iae}, the decoupling hypersurface does not meet either case. 

In this work we propose a new approach for evaluating quantities at LTE, which applies to hypersurfaces with arbitrary 
geometry and, as we will see, features the expected independence on the normal components of the gradients; technically, 
this method relies on an exchange of the order of integrations for the leading order correction 
of the Wigner function, which sheds light on the nature of the correction itself. Although our derivation is focused 
on the spin-1/2 case the method can be readily extended to particles with arbitrary spin. 

The paper is organized as follows. In the first section, we introduce the local equilibrium density operator and define the linear response expansion in terms of thermodynamic gradients. In the second section, we introduce the Wigner function for a free Dirac field and propose a new expansion scheme for its linear response expansion. This scheme allows the integration over the decoupling hypersurface to be performed exactly, at any given order, without resorting to geometrical approximations. In the third section, we apply the resulting formula for the linear response theory of the Wigner function to compute the first-order gradient correction to the spin polarization vector of hadrons. The new formula contains not only gradients of the thermodynamic fields, but also gradients of the normal vector, which are absent in the flat-hypersurface approximation. We finally discuss the physical implications of the new formula.

%=================================================================================================
\subsection*{Notations}
Throughout this paper we use the mostly minus signature convention for the flat metric $g^{\mu\nu}=\mbox{diag}\left(+,-,-,-\right)$. 
We adopt the natural system, $\hbar=k_B=c=1$. The spacial part of a four-vector $k^\mu$ is denoted with the "bold", ${\bf{k}}$, 
the scalar product is denoted with a dot for both four-vectors $k\cdotp p=k_\mu p^\mu$ and three-vectors ${\bf k}\cdot{\bf p}=k_jp^j$. 
Einstein index conventions are assumed, contracted indices are summed all over the possible values, $\mu=0,1,2,3$; $j=1,2,3$. 
The Levi-Civita pseudotensor $\epsilon^{\mu\nu\rho\sigma}$ is chosen so that $\epsilon^{0123}=+1$.

%********************************************************************************
\section{Local thermodynamic equilibrium}
%********************************************************************************

In the calculation of the LTE mean value \eqref{LTEO}, we can replace the four-temperature $\beta_\nu(y)$ and the
reduced chemical potential $\zeta(y)$ in eq. \eqref{LE-density-operator-new} with:
\begin{align*}
 \beta_\nu(y) &= \beta_\nu(x) + \left[ \beta_\nu(y)-\beta_\nu(x) \right] \equiv \beta_\nu(x) + \Delta \beta_\nu(y,x)\;,\\
 \zeta(y) &= \zeta(x)+ [\zeta(y)-\zeta(x)] \equiv \zeta(x)+\Delta\zeta(y,x)\;,
\end{align*}
and rewrite \eqref{LTEO} as:
\begin{align}\label{lteop}
 O_{\rm LE}(x) =&  \frac{1}{Z}  \Tr \left( \exp\left[- \beta(x) \cdot \widehat P {+\zeta(x)\widehat{Q}}  - \int_{\Sigma_{\rm D}} \!\! \di\Sigma_{\mu}(y)\; \left(\widehat{T}^{\mu\nu}(y) 
\Delta\beta_{\nu}(y,x)-\widehat{j}^\mu(y)\Delta\zeta(y,x)\right)\right]\, \widehat{O}(x)\right)\;,
\end{align}
where we have taken out $\beta_\nu(x)$ and $\zeta(x)$ from the integral and used the identity of the four-momentum operator 
$\widehat{P}^\nu = \int_{\Sigma_{\rm D}} \!\! \di\Sigma_{\mu}(y)\; \widehat{T}^{\mu\nu}(y)$ and the charge operator $\widehat{Q}=\int_{\Sigma_{\rm D}} \!\! \di\Sigma_{\mu}(y)\; \widehat{j}^{\mu}(y)$.

In the hydrodynamic limit, with a clear separation between the microscopic scale, associated to
the correlation lengths of the operators, and the scale over which the $\beta$ and $\zeta$ field
significantly varies, the first two terms in the exponent of \eqref{lteop} dominates over the last term which is expected
to provide a correction. We can then use linear response theory to obtain a good approximation of 
the mean value of the operator $\widehat{O}(x)$ at LTE:
$$
O_{\rm LE}(x)=O_{\rm GE}(x)+\Delta O_{\rm LE}(x) +\Delta O_{\rm LE}^{(\zeta)}(x)\;,
$$
where $O_{\rm GE}(x)$ is the global equilibrium mean value calculated with a four-temperature equal to $\beta(x)$ 
and a reduced chemical potential equal to $\zeta(x)$: 
\begin{equation}\label{eq:O-GE}
O_{\rm GE}(x)\equiv\left\langle\widehat{O}(x)\right\rangle_{\rm GE}=\frac{\text{Tr}\left[\widehat{O}(x)\,
\e^{-\widehat{A}(x)}\right]}{\text{Tr}\left[\e^{-\widehat{A}(x)}\right]}\;,
\end{equation}
with $\widehat{A}(x)\equiv\beta(x) \cdot \wP-\zeta(x)\widehat{Q}$. The terms  $\Delta O_{\rm LE}(x)$ and  $\Delta O_{\rm LE}^{(\zeta)}(x)$ are leading corrections from the linear responses to $\Delta\beta$ and $\Delta\zeta$, respectively:
\begin{eqnarray} \label{deltaO}
\Delta O_{\rm LE}(x)&=& -\int_{\Sigma_{\rm D}} \!\! \di \Sigma_{\mu}(y) \; 
\int_{0}^{1} \di z \; \Delta\beta_{\nu}(y,x)\left\langle \widehat{O}(x),\,\e^{- z \widehat{A}(x) } \widehat{T}^{\mu\nu}(y) 
\e^{z\widehat{A}(x)} \right\rangle _{c,\text{GE}}\;, \nonumber\\ 
\Delta O_{\rm LE}^{(\zeta)}(x)&=&\int_{\Sigma_{\rm D}} \!\! \di \Sigma_{\mu}(y) \; 
\int_{0}^{1} \di z \; \Delta\zeta(y,x)\left\langle \widehat{O}(x),\,\e^{- z \widehat{A}(x) } \widehat{j}^{\mu}(y) 
\e^{z\widehat{A}(x)} \right\rangle _{c,\text{GE}}\;.
\end{eqnarray}
 In the above equations, $\left\langle\cdots\right\rangle_{c,\text{GE}}$ denotes the connected mean value at GE,
i.e:
\begin{equation*}
    \left\langle\widehat{O}_1(x),\,\widehat{O}_2(y)\right\rangle_{c,\rm GE}=\left\langle\widehat{O}_1(x)\widehat{O}_2(y)\right\rangle_{c,\rm GE}-\left\langle\widehat{O}_1(x)\right\rangle_{\rm GE}\left\langle\,\widehat{O}_2(y)\right\rangle_{\rm GE}\;.
\end{equation*}
%
%---------------------------------------------------------------------------------
\begin{figure}
\includegraphics[width=8cm]{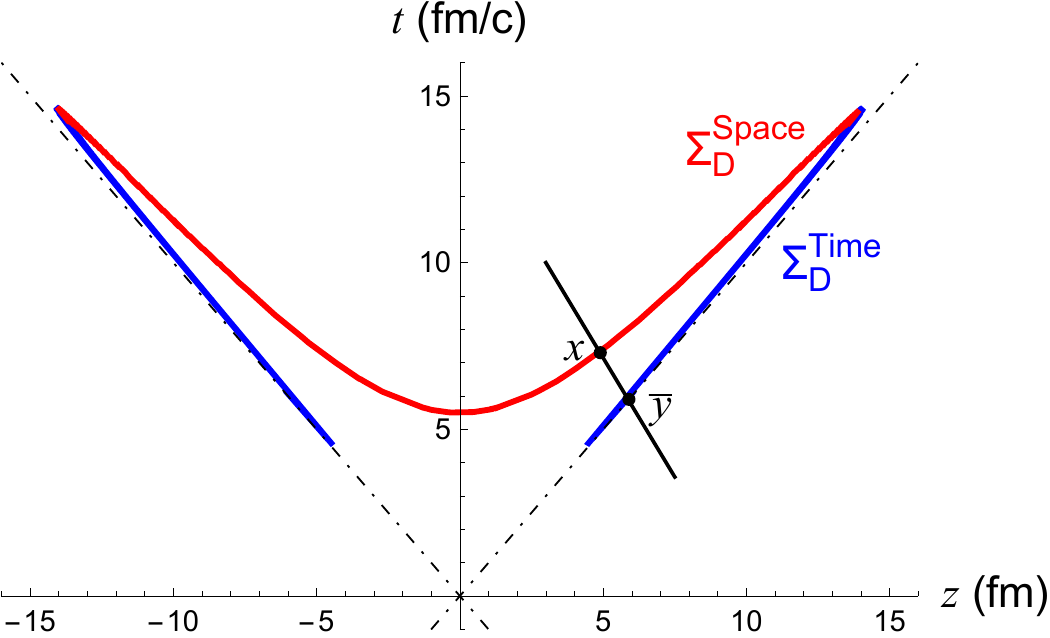}
\caption{\label{fig:Freeze-out} A slice of the decoupling hypersurface $\Sigma_{\rm D}$
(red and blue lines) on the transverse plane $x=y=0$, obtained by 3+1D viscous hydrodynamic
simulations with the CLVisc model at $\sqrt{s_{\text{NN}}}=27$ GeV \citep{Pang:2018zzo,Wu:2021fjf,Wu:2022mkr}.
The black solid line represents the worldline of a particle, which intersects $\Sigma_{\rm D}$ at 
two points: one on the spacelike branch and the other one on the timelike branch.}
\end{figure}
%----------------------------------------------------------------------------------

%********************************************************************************
\section{Wigner function for spin-1/2 particles}
%********************************************************************************

%-----------------------------
\subsection{Wigner operator}
%-----------------------------
The covariant Wigner function $W(x,p)$ is a quantum generalization of the classical phase 
space distribution \cite{DeGroot:1980dk,Heinz:1983nx,Elze:1986qd,Vasak:1987um}; all physical 
observables can be expressed with suitable integrals of the Wigner function. For free 
spin-1/2 particles, it is defined as the mean value of the Wigner operator:
\begin{equation}\label{Wigner-operator}
\widehat{W}_{ab}^+(x,p)\equiv\theta(p^0)\!\int\frac{\di^{4}y}{(2\pi)^{4}}\e^{-\ii p\cdot y}
\overline{\Psi}_{b}\left(x+\frac{y}{2}\right)\Psi_a\left(x-\frac{y}{2}\right),
\end{equation}
where $\Psi$ denotes the Dirac field and $a$, $b$ are spinor indices. The step function $\theta(p^0)$ 
ensures that only particle contributions are included (for antiparticles the procedure is alike). 
The Dirac field can be expanded in terms of plane waves:
\begin{equation}\label{eq:quantized-psi}
\Psi(x)=\frac{1}{\sqrt{(2\pi)^{3}}}\sum_s\int\frac{\di^{3}{\bf \p}}{2\varepsilon} 
\left(u_{s}(p) \widehat{a}_{s}(p)\e^{-ip\cdot x} + v_s(p) \widehat{b}_s^\dagger \e^{\ii p \cdot x} \right)\;,
\end{equation}
where $\varepsilon =\sqrt{{\bf p}^{2}+m^{2}}$ denotes the on-shell energy and $\widehat{a}_s(p_-)$ is the annihilation operator for a particle 
with momentum $p_-$ (note that $p_\pm$ is on-shell) and spin component $s$ and the $u$'s are the
familiar spinors solution of the free Dirac equation. The creation and annihilation operators 
obey the covariant anti-commutation relation:
\begin{equation*}
\left\{ \widehat{a}_{s}(p),\widehat{a}_{s^{\prime}}^{\dagger}(p^{\prime})\right\} =
2\varepsilon_{{\bf p}}\delta^{3}({\bf p}-{\bf p}^{\prime})\delta_{ss^{\prime}}\;,
\end{equation*}
and likewise for antiparticles. Replacing the integral measure in eq. \eqref{eq:quantized-psi} with:
\begin{equation*}
\int\frac{\di^3 \p}{2\varepsilon}\rightarrow \int \di^4 p\; \delta(p^{2}-m^{2})\theta(p^{0})\;,
\end{equation*}
and substituting the Dirac field into \eqref{Wigner-operator}, the Wigner operator for the particle
part turns out to be:
\begin{eqnarray}\label{eq:Wigner-operator}
 \widehat{W}^{+}(x,p) & = & \frac{1}{(2\pi)^3} \sum_{r,s} \int \di^4 p_1 \, \di^4 p_2 \; 
 \delta(p_1^{2}-m^{2})\theta(p_1^{0}) \delta(p_2^2-m^{2})\theta(p_2^{0}) \,
 \delta\left( p-\frac{p_1+p_2}{2}\right) \nonumber \\
&& \times \e^{\ii (p_1-p_2) \cdot x} \widehat{a}_{r}^{\dagger}(p_1)\widehat{a}_{s}(p_2) 
u_{s}(p_2)\bar{u}_{r}(p_1)\;.
\end{eqnarray}
One can now change the integration variables:
\be\label{varchange}
 P \equiv \frac{p_1+p_2}{2}\;, \qquad q \equiv p_1-p_2\;, \qquad\implies\qquad p_1 = P + q/2\;, \qquad p_2 = P - q/2\;,
\ee
so that:
$$
  \di^4 p_1 \; \di^4 p_2 = \di^4 P \; \di^4 q\;,
$$
and 
$$
\delta(p_1^{2}-m^{2})\delta(p_2^{2}-m^{2})=\frac{1}{2}\delta\left(P^{2}+\frac{q^{2}}{4}-m^{2}\right)
\delta(P\cdot q)\;.
$$
Integrating in $\di^4 P$ is now straightforward because of the $\delta^4 (p - P)$ in the equation
\eqref{eq:Wigner-operator} and, after replacing all variables according to the \eqref{varchange} one 
readily gets 
the following integral expression:
\begin{eqnarray}\label{Wigner-operator2}
\widehat{W}^+(x,p) &=& \frac{1}{2(2\pi)^3}\sum_{r,s} \int \di^{4} q \; \e^{\ii q\cdot x}
\theta(p_+^0)\theta(p_-^0) \delta(p \cdot q)  \nonumber \\
 && \hspace{-1.5cm}\times \delta \left( p^2 + \frac{q^2}{4} - m^2 \right) 
 \wad_r (p_+) \wa_s(p_-) u_s (p_-) \bar u_r(p_+)\;,
\end{eqnarray}
where $p_\pm\equiv p\pm q/2$ are restricted on the mass shell.

%-------------------------------------------------------------------
\subsection{Leading corrections to Wigner function}
%-------------------------------------------------------------------

We now focus on the leading corrections to the Wigner function in the linear response theory, which 
can be derived by replacing $\widehat{O}$ in eq. \eqref{deltaO} by the Wigner operator 
$\widehat{W}^+(x,p)$. Plugging the equation \eqref{Wigner-operator2} into the first line in \eqref{deltaO} we obtain:
\begin{eqnarray}\label{eq:delta-W-LE}
\Delta W_{\text{LE}}^{+}(x,p) & = & -\frac{1}{2(2\pi)^{3}}\sum_{r,s}\int 
\di^{4}q \; \theta(p_{+}^{0})\theta(p_{-}^{0}) \e^{\ii q\cdot x} \,\delta(p\cdot q)\int_{\Sigma_{\rm D}}\!\! 
\di \Sigma_{\mu}(y) \int_{0}^{1}\di z \; \Delta\beta_{\nu}(y,x) \, \delta\left(p^{2}+\frac{q^{2}}{4}-m^{2}\right) 
\nonumber \\
&& \times \left\langle \widehat{a}_{r}^{\dagger}(p_{+})\widehat{a}_{s}(p_{-}), \e^{-z\widehat{A}(x)} 
\widehat{T}^{\mu\nu}(y) \e^{z\widehat{A}(x)}\right\rangle_{c,\text{GE}}u_{s}(p_{-})\bar{u}_{r}(p_{+})\;.
\end{eqnarray}
Since the four-momentum operator $\widehat{P}^\mu$ is the generator of spacetime translation, we have:
\begin{equation*}
    \wT^{\mu\nu}(y)=\e^{\ii\wP\cdot y}\wT^{\mu\nu}(0)\e^{-\ii\wP\cdot y}\;,
\end{equation*}
and taking into account that $\widehat{A}(x)=\beta(x) \cdot \wP-\zeta(x)\widehat{Q}$ commutes with $\widehat{P}^\mu$:
\begin{equation}\label{eq:aaTmunu}
\left\langle \widehat{a}_{r}^{\dagger}(p_{+})\widehat{a}_{s}(p_{-}), \e^{-z\widehat{A}(x)} 
\widehat{T}^{\mu\nu}(y) \e^{z\widehat{A}(x)}\right\rangle_{c,\text{GE}}=\left\langle e^{-\widehat{C}(x)}\widehat{a}_{r}^{\dagger}(p_{+})\widehat{a}_{s}(p_{-})e^{\widehat{C}(x)}, 
\widehat{T}^{\mu\nu}(0)\right\rangle_{c,\text{GE}}\;,
\end{equation}
where $\widehat{C}(x)=-z[\beta(x)\cdot \wP -\zeta(x)\wQ]+iy\cdot \wP$. Since $\wad(p_\pm)$,$\wa(p_\pm)$ are the 
creation/annihilation operators of eigen-momentum states, they transform under translation as:
\begin{align*}
        \e^{\ii\wP\cdot y}\wad(p)\e^{-\ii\wP\cdot y}&=\e^{\ii p\cdot y}\wad(p)\;,\\
        \e^{\ii\wP\cdot y}\wa(p')\e^{-\ii\wP\cdot y}&=\e^{-\ii p'\cdot y}\wa(p')\;,
\end{align*}
so that:
\begin{equation*}
 \e^{-\widehat{C}(x)} \wad(k_+) \wa(k_-) \e^{\widehat{C}(x)} 
=\e^{z\beta(x)\cdot q-\ii y\cdot q }\wad(k_+)\wa(k_-)\;.
\end{equation*}
Plugging eq. \eqref{eq:aaTmunu} into eq. \eqref{eq:delta-W-LE} and completing the integral over $z$ leads to:
\begin{align}\label{Wigner-off-GTE}
\Delta W_{\rm LE}^+(x,p) &=\frac{1}{2(2\pi)^3}\sum_{r,s} \int_{\Sigma_{\rm D}} \!\! 
\di \Sigma_{\mu}(y)\Delta \beta_\nu(y,x) \int \di^{4}q\; \e^{\ii q\cdot (x-y)} 
\delta(p\cdot q)\, \frac{1 -\e^{\beta(x)\cdot q}}{\beta(x) \cdot q} 
\nonumber \\
& \times  \delta \left( p^2 + \frac{q^2}{4} - m^2 \right)  \theta(p_+^0)\theta(p_-^0)\,  u_s (p_-)\bar u_r(p_+) \left\langle \wad_r (p_+) \wa_s(p_-), \wT^{\mu\nu}(0) 
\right\rangle_{c,{\rm GE}} \,.
\end{align}
The linear response to $\Delta\zeta$, $\Delta W^{+(\zeta)}_{\rm LE}$, is obtained by replacing $\Delta\beta_\nu(y,x)$ 
with $\Delta \zeta(y,x)$ and $\wT^{\mu\nu}(0)$ with $-\wj^\mu(0)$.

%-------------------------------------------------------
\subsection{Integration over the decoupling hypersurface} 
\label{sec:hypersurface-int}
%-------------------------------------------------------

In ref. \cite{Becattini:2021suc}, the integration over $\Sigma_{\rm D}$ in eq. \eqref{Wigner-off-GTE} was carried 
out first, introducing a non-trivial geometric assumption, 
followed by the integration in $q$. In fact, it is more convenient to carry out the integration in $q$ first because it
does not require any special assumption other than the usual hydrodynamic limit. This reverted order of integration
eventually does not imply special geometric assumptions, as has been mentioned in the Introduction, and it generates
a nice expression, as we will see.

The formula \eqref{Wigner-off-GTE} can be rewritten in a way which makes it apparent the
effect of the hydrodynamic limit:
\be\label{deltaWig}
\Delta W_{\rm LE}^+(x,p) =\frac{1}{(2\pi)^3} 
\int \di^4 q\; \delta (p \cdot q) G^{\mu\nu}(q) F_{\mu\nu}(q)\;,
\ee
where:
\be\label{Gfunct}
G^{\mu\nu}(q) = \frac{1}{2} \delta \left( p^2 + \frac{q^2}{4} - m^2 \right) 
\frac{1 - \e^{\beta(x)\cdot q}}{\beta(x) \cdot q} \theta(p_+^0)\theta(p_-^0) 
\sum_{r,s} u_s (p_-) \bar u_r(p_+)\left\langle \wad_r (p_+) \wa_s(p_-),\, 
\wT^{\mu\nu}(0) \right\rangle_{c,{\rm GE}} \;,
\ee
and
$$
 F_{\mu\nu}(q) = \int_{\Sigma_{\rm D}} \!\!  \di \Sigma_{\mu}(y) \; 
\e^{-iq\cdot(y-x)} \Delta \beta_\nu(y,x)\;.
$$
In the hydrodynamic limit, $\Delta\beta_{\nu}$ is a slowly varying function in space and time implying that 
the function $F_{\mu\nu}$, which is a Fourier transform in the variable $q$ integrated in the variable $y$, 
is a peaked function about $q^{\mu}=0$ (also taking into account that $q\cdot p =0$). A good approximation of
the integral \eqref{deltaWig} should be then obtained by a small-$q$ expansion of the function $G^{\mu\nu}(q)$ 
in the integral \eqref{deltaWig}:
\begin{equation}\label{Gqexpans}
G^{\mu\nu}(q)=\sum_{N=0}^{\infty}\frac{1}{N!}\left.\left[\partial^{q}_{\nu_{1}}
\partial^q_{\nu_2}\cdots\partial^{q}_{\nu_{N}}G^{\mu\nu}(q)\right]\right|_{q^{\mu}=0}q^{\nu_1}q^{\nu_2}\cdots q^{\nu_N}\;.
\end{equation}
Indeed, as it will become clear later in equation \eqref{eq:wignerfinal}, this expansion in $q$ about $q=0$
generates an expansion of the Wigner function in the gradients of the thermo-hydrodynamic field $\beta$. Because
of the presence of the $\delta$ and $\theta$ distributions in the function \eqref{Gfunct}, the expansion
\eqref{Gqexpans} should be interpreted in a distributional sense, i.e. meaningful only upon eventual integration
in $\di k^0$ as specified in the equation \eqref{spin-polarization-Wigner} later on.

Plugging the \eqref{Gqexpans} into the \eqref{deltaWig}, the leading order correction to the Wigner function 
becomes:
\be\label{eq:W-LTE-In}
\Delta W_{\rm LE}^+(x,p) = \frac{1}{(2\pi)^3}
\sum_{N=0}^{\infty}\frac{1}{N!}\int_{\Sigma_{\rm D}} \!\! \di \Sigma_{\mu}(y) \,
I_{n}^{\nu_{1}\nu_{2}\cdots\nu_{N}}(y-x) \Delta\beta_\nu(y,x) \left[\partial_{\nu_{1}}^{q}
\partial_{\nu_{2}}^{q}\cdots\partial_{\nu_{}}^{q} G^{\mu\nu}(q)\right]\Big|_{q=0}\;,
\ee
where integrals over $q$ are encoded in the rank-$N$ function 
\be\label{eq:I-n-final}
I_{N}^{\nu_{1}\nu_{2}\cdots\nu_{n}}(y-x)\equiv\int \di^{4}q\,\delta(p\cdot q)
\e^{-\ii q\cdot(y-x)}q^{\nu_{1}}\cdots q^{\nu_{N}} =(2\pi)^{3}\frac{(-\ii)^{N}}{|p^{0}|}
\partial_x^{\nu_1}\cdots \partial_x^{\nu_N}
 \delta^{3}\left({\bf y}-{\bf x}-\frac{\bf p}{p^{0}}(y^{0}-x^{0})\right)\;.
\ee
Plugging the eq. \eqref{eq:I-n-final} into the \eqref{eq:W-LTE-In} and repeatedly using the Leibniz rule 
(as shown in Appendix \ref{Appendix:A}) the equation \eqref{eq:W-LTE-In} can be transformed into:
\begin{eqnarray}\label{eq:delta-W-2}
\Delta W_{\text{LE}}^+(x,p)& = & \frac{1}{|p^0|} \sum_{N=0}^{\infty} \frac{(-i)^N}{N!}
\left.\left[\partial^{q}_{\nu_{1}}\cdots\partial^{q}_{\nu_{N}}G^{\mu\nu}(q)\right]\right|_{q=0}
\nonumber \\
&\times&\sum_{M=0}^N \frac{N!(-1)^M}{M!(N-M)!}\partial^{\nu_{M+1}}_{x}\cdots\partial^{\nu_{N}}_{x}
\int_{\Sigma_{\rm D}}\!\! \di \Sigma_{\mu}(y) 
\left[\partial^{\nu_{1}}_{x}\cdots\partial^{\nu_{M}}_{x}\Delta\beta_{\nu}(y,x)\right]
 \delta^3 \left({\bf y}-{\bf x}-\frac{\bf p}{p^0}(y^0-x^0)\right)\;.
\end{eqnarray}
In general, $\Sigma_{\rm D}$ can be a topological complex hypersurface, with both space-like and time-like parts. 
Notably, for a given spatial position ${\bf x}$ there might be multiple $x^0$ times belonging to $\Sigma_{\rm D}$ (see fig. \ref{fig:Freeze-out}). Nevertheless, the hypersurface can be split into single-valued branches $x^0=f_k({\bf x})$
and the integration in \eqref{eq:delta-W-2} can be decomposed into the sum of integrals over those branches. For 
any arbitrary function $\Theta(x,y)$ the integration in $y$ in the eq. \eqref{eq:delta-W-2} yields (assuming that 
$p\cdot \sigma \ne 0$):
\begin{eqnarray}\label{deltas}
&& \int_{\Sigma_{\rm D}} \!\! \di \Sigma_{\mu}(y) \; \delta^{3}\left({\bf y}-{\bf x}-\frac{\bf p}{p^{0}}
(y^{0}-x^{0})\right) \Theta(y,x) \nonumber \\
&& = \sum_{k} s_k \int_{\Sigma_k} \!\! \di^3 {\rm y} \; \sigma_{\mu}(y)\;  
\delta^{3}\left({\bf y}-{\bf x}-\frac{\bf p}{p^{0}}(y^{0}-x^{0})\right) \Theta(y,x) \nonumber \\
&& = \sum_{k,i} s_k \sigma_{\mu}(\bar y_{k,i})\; \frac{|p^0|}{|p\cdot\sigma(\bar y_{k,i})|} 
\Theta(\bar y_{k,i},x)\;,
\end{eqnarray} 
where $\sigma$ is the vector perpendicular to the hypersurface:
\begin{equation}\label{eq:normal-vector}
\sigma_{\mu}(x)=\left(1,-\partial f_k/\partial{\bf x}\right)\;,
\end{equation}
and $\bar{\bf y}_{k,i}$ is the $i$-th solution of the equation ${\bf y} = {\bf x}+({\bf p}/p^0)
(f_k({\bf y})-x^{0}({\bf x}))$; $s_k$ in eq. \eqref{deltas} is a sign, which is +1 if $\sigma_\mu$ 
has the same direction of the outward pointing vector $\di \Sigma_\mu$ and $-1$ otherwise.
For any given ${\bf x}$ and $x^0$ lying on $\Sigma_{\rm D}$, the solutions are found by 
intersecting the world-line of a particle off the mass-shell (being $p^2 \ne m^2$) moving 
with velocity ${\bf p}/p^0$:
\be\label{worldline}
 {\bf y} = {\bf x}+\frac{\bf p}{p^{0}}(y^0-x^0)\;,
\ee
with all branches of the hypersurface $\Sigma_{\rm D}$. This is depicted in fig. \ref{fig:Freeze-out};
there is at least one solution which is $y=x$, but there might be more if the topology
of the hypersurface makes it possible. We can then replace the sum over $k,i$ in the 
equation \eqref{deltas} with the sum over the $p$-dependent intersections $\bar y(x,p)$ for a 
given $x$.

Plugging the eq. \eqref{deltas} into the eq. \eqref{eq:delta-W-2}, one obtains:
\begin{eqnarray}\label{eq:delta-W-3}
\Delta W_{\text{LE}}^{+}(x,p) & = & \sum_{N=0}^{\infty}\sum_{\bar{y}(x,p)}\frac{(-i)^{N}}{N!}
\left.\left[\partial^{q}_{\nu_{1}}\ldots\partial^{q}_{\nu_{N}}G^{\mu\nu}(q)\right]\right|_{q=0}\nonumber \\
 && \times\sum_{M=0}^{N}\frac{N!(-1)^{M}}{M!(N-M)!}\di^{\nu_{M+1}}_{x}\ldots\di^{\nu_{N}}_{x}
 \left[\left.\frac{s_k \sigma_{\mu}(y)}{|p\cdot\sigma(y)|}\partial^{\nu_{1}}_{x}\ldots\partial^{\nu_{M}}_{x}
 \Delta\beta_{\nu}(y,x)\right|_{y=\bar{y}}\right]\;,
\end{eqnarray}
where the sum over $\bar y(x,p)$ runs over all the 
intersections of the particle world-line with the hypersurface for a chosen space-time point $x$, 
including the trivial solution $y=x$. In the equation \eqref{eq:delta-W-3} we have introduced the total 
derivative:
$$
\di^\mu_x = \frac{\di}{\di x_\mu}\;,
$$
to emphasize the difference between the derivative acting on the function {\em before} setting 
$y=\bar{y}(x,p)$ (that is $\partial_x$) and the derivative acting on the function {\em after}
setting $y=\bar{y}(x,p)$. In general, using the chain rule for an arbitrary function $g(x,\bar{y}(x,p))$ we have:
\begin{equation}\label{chain-rule}
\frac{\di}{\di x^\mu} g(x,\bar{y}(x,p))= \frac{\partial}{\partial x^\mu} g(x,\bar{y}(x,p))+ 
\frac{\partial \bar{y}^\nu(x,p)}{\partial x^\mu} \frac{\partial g(x,y)}{\partial y^\nu}\Bigg|_{y=\bar{y}(x,p)}\;.
\end{equation}
It can be shown (see Appendix \ref{Appendix:A} for the detailed calculation) that: 
\begin{equation}\label{eq:dydx}
 \frac{\partial\bar{y}^{\nu}(x,p)}{\partial x^{\mu}}=\delta_{\mu}^{\nu}-\frac{p^{\nu}
 \sigma_{\mu}(\bar{y})}{p\cdot\sigma(\bar{y})} \equiv \Delta_{\mu}^{\;\nu}(\bar{y})\;.
\end{equation}
Since $n_\mu= s_k\sigma_\mu/\sqrt{|\sigma\cdot\sigma|}$ is the outward pointing unit vector normal 
to the hypersurface $\Sigma_{\rm D}$, the tensor $\Delta^{\nu\rho}$ can be also expressed as: 
\be\label{operatorDelta}
 \Delta^{\nu\rho}(\bar{y}) = g^{\nu\rho} - \frac{n^\nu(\bar{y}) p^\rho}
 {p \cdot n(\bar{y})}\;.
\ee
Hence, by using the \eqref{chain-rule} and the above definitions, the eq. \eqref{eq:delta-W-3} can be rewritten 
as:
\begin{eqnarray}\label{eq:wignerquasifinal}
\Delta W_{\text{LE}}^{+}(x,p) & = &\sum_{N=0}^{\infty} \sum_{\bar{y}(x,p)} \frac{(-i)^{N}}{N!}
\left.\left[\partial^{q}_{\nu_{1}}\cdots\partial^{q}_{\nu_{N}}G^{\mu\nu}(q)\right]\right|_{q=0} 
\sum_{M=0}^{N}\frac{N!(-1)^{M}}{M!(N-M)!}\left[\partial^{\nu_{M+1}}_{x}+
\Delta^{\nu_{M+1}\rho_{M+1}}(\bar{y})\partial_{\rho_{M+1}}^{y}\right]\nonumber \\
 && \times\left.\cdots\left[\partial^{\nu_{N}}_{x}+\Delta^{\nu_N\rho_{N}}(\bar{y})
 \partial_{\rho_{N}}^{y}\right]\partial^{\nu_{1}}_{x}\cdots\partial^{\nu_{M}}_{x}
 \frac{n_{\mu}(y)}{|p\cdot n(y)|}\Delta\beta_{\nu}(y,x)\right|_{y=\bar{y}}\;.
\end{eqnarray}
In this result, $\bar{y}$ is a function of $x$ and therefore $\partial_x^\nu$ does not commute with $\Delta^{\nu\rho}(\bar{y})\partial_\rho^y$. With the help of 
\begin{equation*}
\partial_x^\nu\Delta^{\rho\alpha}(\bar{y})=\frac{\partial y^\beta}{\partial x_\nu}\partial^{\bar y}_\beta\Delta^{\rho\alpha}(\bar{y})=\left.\Delta^{\nu\beta}(y)\partial_\beta^y\Delta^{\rho\alpha}(y)\right|_{y=\bar{y}}\,,
\end{equation*}
which can be obtained by using the chain rule and the eq. (\ref{eq:dydx}), we can replace all $\Delta(\bar{y})$ in Eq. (29) by $\Delta(y)$ and using the Binomial theorem:
$$
  D^{(N)} = (-\partial_x + \partial_x + D)^{(N)} = \sum_{M=0}^N \binom{N}{M} (-\partial_x)^{(M)} (\partial_x+D)^{(N-M)}\;.
$$
The equation \eqref{eq:wignerquasifinal} can be recast into a compact form:
\be\label{eq:wignerfinal}
\Delta W_{\text{LE}}^+(x,p) =\sum_{N=0}^{\infty}\sum_{\bar{y}(x,p)}\frac{1}{N!}
D^N_y(y) \left[ G^{\mu\nu}(q) \frac{n_{\mu}(y)}{|p\cdot n(y)|}
\Delta\beta_{\nu}(y,x) \right] \Bigg|_{q=0,y=\bar y(x,p)}\;.
\ee
The operator $D_y$ in eq. \eqref{eq:wignerfinal} is defined as:
\be\label{eq:operator-D-q-y}
 D_y(y) \equiv - \ii \Delta^{\nu\rho}(y) \partial^y_\rho \partial^q_\nu
\ee
and the derivative can act on the operator $D_y(y)$ on its right-hand-side.

At any given $N$, the right hand side of eq. \eqref{eq:wignerfinal} contains derivatives of $N$-th order in space
and momentum $q$, so the $q$ expansion is equivalent to a gradient expansion. The formula includes gradients of 
$\beta_\nu$ as well as gradients of the normal vector $n$ to the hypersurface; the latter are obviously absent if 
the hypersurface is a hyperplane, as it is tacitly assumed in many calculations in thermal field theory, and yet
for a curved hypersurface their contribution may not be negligible. Note that the $N-$th term of the series 
contains mixed terms made by combinations of derivative of the thermodynamic fields and curvature term are present.

A crucial features of the equation \eqref{eq:operator-D-q-y} is the absence of the normal component of the spacial
gradients. By decomposing the gradient into a normal component along $n$ and a tangential component to the hypersurface
perpendicular to $n$:
\be\label{ddecomp}
  \partial^y_\rho = n_\rho (n \cdot \partial^y) + \partial_{T\rho}^y \equiv n_\rho D + \partial_{T\rho}^y\;,
\ee
it turns out that only the tangential component survives:
$$
D_y(y)= -\ii \Delta^{\nu\rho}(y) \partial_{\rho}^{y}\partial_{\nu}^{q} = 
\left( g^{\nu\rho} - \frac{n^\nu(y) p^\rho}
 {p \cdot n(y)}\right) \left( n_\rho D + \partial_{T\rho}^y \right)\partial_{\nu}^{q}
 = -\ii \Delta^{\nu\rho}(y) \partial_{T\rho}^{y}\partial_{\nu}^{q}\;.
$$
This result is precisely what one should have expected in view of the general conclusion about the independence
of the LTE mean values of the thermodynamic fields outside the hypersurface $\Sigma_{\rm D}$ that was discussed in the
Introduction section.

The calculation of the correction $\Delta W_{\text{LE}}^{+(\zeta)}$ follows the same steps as the previous one
and one eventually gets:
\be\label{eq:wignerfinal-zeta}
\Delta W_{\text{LE}}^{+(\zeta)}(x,p) =\sum_{N=0}^{\infty}\sum_{\bar{y}(x,p)}\frac{1}{N!}
D^N_y(y) \left[ H^{\mu}(q) \frac{n_{\mu}(y)}{|p\cdot n(y)|}
\Delta\zeta(y,x) \right] \Bigg|_{q=0,y=\bar y(x,p)}\;,
\ee
where 
\be\label{Hfunct}
H^{\mu}(q) = -\frac{1}{2} \delta \left( p^2 + \frac{q^2}{4} - m^2 \right) 
\frac{1 - \e^{\beta(x)\cdot q}}{\beta(x) \cdot q} \theta(p_+^0)\theta(p_-^0) 
\sum_{r,s} u_s (p_-) \bar u_r(p_+)\left\langle \wad_r (p_+) \wa_s(p_-), 
\wj^{\mu}(0) \right\rangle_{c,{\rm GE}}\;. 
\ee
%

%------------------------------------------------------------
\subsection{Discussion}
%------------------------------------------------------------

It is somewhat surprising that the Wigner function at the point $x$ receives contributions from the gradients of 
the fields $\beta$ and $\zeta$ evaluated at points $\bar y(x,p)$
which can be relatively distant from $x$. These points lie on the world-line \eqref{worldline} 
but this by no means implies that there is an actual particle travelling from $\bar y(x,p)$ 
to $x$; its possible physical origin is discussed towards the end of the paper. The 
mathematical origin of this long-distance contribution can be understood rewriting the 
equation \eqref{Wigner-off-GTE} as:
\be\label{deltawig2}
\Delta W_{\rm LE}^+(x,p) =\frac{1}{(2\pi)^3} 
\int_{\Sigma_{\rm D}} \!\! \di \Sigma_{\mu}(y) \Delta \beta_\nu(y,x)
\left[ \int \di^{4}q\; \delta(p\cdot q) \e^{\ii q\cdot (x-y)} G^{\mu\nu}(q) \right]\;. 
\ee
In order for $y=x$ to be the only dominant contribution of the integral \eqref{deltawig2}, 
the function of $y-x$ between square brackets should be peaked around zero, but because of 
the presence of $\delta(p\cdot q)$, this function is constant through the whole worldline 
\eqref{worldline}, so that points other than $x$ may give a large contribution 
to the integral.

An important question is about the number of terms which should be computed in the series \eqref{eq:wignerfinal}
and \eqref{eq:wignerfinal-zeta} in order to obtain a satisfactory approximation. The speed of convergence of these 
series apparently depends on the ratio between two lengths; first, the correlation length $l_c$ associated to the 
function $G^{\mu\nu}$ ($H^\mu$), that is $ \left|\partial_q G^{\mu\nu}(q)\right|/\left|G^{\mu\nu}(q)\right|$ 
($\left|\partial_q H^{\mu}(q)\right|/\left|H^{\mu}(q)\right|$), which depends on the microscopic 
scales of the quantum field and the inverse temperature. 
Second, the characteristic variation lengths of the thermo-hydrodynamic field $\beta$, $L_H = | \beta |/ \partial \beta|$ 
($| \zeta |/ \partial \zeta|$), and that of the normal vector $n$, $L_G = | n |/ \partial n|$ which is 
a purely geometric quantity. So, the series will rapidly converge if:
\begin{equation}\label{converge-requirement}
l_{c}/L_{H}\ll1\;,\qquad l_{c}/L_{G}\ll1\;.
\end{equation}
Such conditions are believed to be generally satisfied in high-energy heavy-ion collisions, 
yet the second one has not been carefully examined thus far. 
Here we provide a simple quantitative estimate for a hypersurface with constant proper time $\tau$. The normal vector is approximated 
as $n^\mu=(\cosh\xi,\,0,\,0,\,\sinh\xi)$, with $\xi$ being the spacetime rapidity. The typical scale of its curvature is $L_G\sim\tau\cosh^{-1}\xi$. For most central Au+Au collisions at $\sqrt{s_{\rm NN}}=200$ GeV, $\tau\sim 10$ fm/c 
\citep{Schafer:2021csj}, corresponding to $6.5\,\text{fm}\leq L_G\leq 10\,\text{fm}$ within the interval $|\xi|<1$, which is 
much larger than the typical microscopic scale $l_c\sim 1/T\sim 1.2$ fm when $T=160$ MeV. In the peripheral region with a larger 
$\xi$, the hypersurface becomes more curved with a larger $l_c/L_G$, and it is not obvious whether the contribution of 
derivatives of the normal vector at all orders with fixed order in the derivatives of $\Delta \beta$ may be important. 

%********************************************************************************
\section{Spin polarization}
%********************************************************************************

With our formula \eqref{eq:wignerfinal} one can in principle calculate gradient corrections at LTE to all measurable quantities, 
notably particle spectra and spin polarization. We focus on the spin polarization of spin-$1/2$ fermions,
which is one of the most studied cases and also plays a crucial role for studying vector meson's spin alignment 
\cite{Liang:2004xn,Chen:2023hnb,Chen:2024afy}, hyperon's spin correlation \cite{STAR:2025njp,Chen:2024hki,Sheng:2025puj}, 
and hypernuclei's spin polarization \cite{Sun:2025oib,Liu:2025kpp}.

With the Wigner function, one can readily calculate the mean spin polarization using the formula \cite{Becattini:2020sww}:
\begin{equation}\label{spin-polarization-Wigner}
S^{\mu}(p)=\frac{1}{2} \frac{\int \di p^0\int_{\Sigma_{\rm D}} \!\! \di \Sigma\cdot p\,\text{tr}\left(\gamma^{\mu}\gamma^{5}W^+
(x,p)\right)}{\int \di p^0 \int_{\Sigma_{\rm D}} \!\! \di \Sigma\cdot p\,\text{tr}(W^+(x,p))}\;,
\end{equation} 
where the leading contribution to the denominator arises from $W^+_\text{GE}$. With the definition \eqref{eq:O-GE}, one 
can calculate the following mean value
$$
\langle \wad_r(p) a_s(p') \rangle_{\rm GE} = 
   2 \varepsilon \, \delta^3({\bf p}-{\bf p}')\delta_{rs} n_F(x,p)\;,
$$
where $\varepsilon=\sqrt{{\bf p}^2+m^2}$ is the on-shell energy and $n_F(x,p)=[1+e^{\beta(x)\cdot p-\zeta}]^{-1}$ denotes the 
Fermi-Dirac distribution. Plugging this equation into the definition of the Wigner operator for the particles
eq. \eqref{Wigner-operator2} the following expression is obtained:
\begin{equation}\label{Wigner-GE}
  W_{\rm GE}^+(x,p)=\frac{1}{(2\pi)^3}\delta(p^2-m^2)\theta(p^0)(\slashed{p}+m)n_F(x,p)\;,
\end{equation}
whose trace (in the 4-dimensional spinor space) reads
\begin{equation}\label{scalar-component}
\text{tr}\left( W_{\rm GE}^+(x,p) \right) =\frac{4m}{(2\pi)^3}\delta(p^2-m^2)
\theta(p^0)n_F(x,p)\;.
\end{equation}
On the other hand, since $\text{tr}\left(\gamma^\mu\gamma^5 W_{\rm GE}^+(x,p)\right)=0$, the leading contribution to the 
numerator in eq. \eqref{spin-polarization-Wigner} arises from $\Delta W_{\rm LE}^+$. 

For a free Dirac field, the Belinfante symmetric energy-momentum tensor and the current operator are given by
\begin{align}
\widehat{T}^{\mu\nu}(x)&=\frac{i}{4}\bar{\Psi}(x)\left[\gamma^{\mu}\left(\overrightarrow{\partial}^{\nu}-\overleftarrow{\partial}^{\nu}\right)+\gamma^{\nu}\left(\overrightarrow{\partial}^{\mu}-
\overleftarrow{\partial}^{\mu}\right)\right]\Psi(x)\;, \nonumber\\
\widehat{j}^\mu(x)&=\bar\Psi(x)\gamma^\mu\Psi(x)\;,
\end{align}
Using the quantized field operator in eq. \eqref{eq:quantized-psi}, the operators $\widehat{T}^{\mu\nu}(0)$ and 
$\widehat{j}^\mu(0)$ turn out to be:
\begin{align}
\widehat{T}^{\mu\nu}(0) =& \frac{1}{2(2\pi)^{3}} \sum_{r,s} \int \di^{4}p_{+}\di^{4}p_{-} \;
\delta(p_{+}^{2}-m^{2})\delta(p_{-}^{2}-m^{2})\bar{u}_{s}(p_{-})(\gamma^{\mu}p^{\nu}+\gamma^{\nu}p^{\mu})
u_{r}(p_{+})\widehat{a}_{s}^{\dagger}(p_{-})\widehat{a}_{r}(p_{+}) \theta(p^0_+)\theta(p^0_-)\;,\nonumber\\
\widehat{j}^{\mu}(0) =& \frac{1}{(2\pi)^{3}} \sum_{r,s} \int \di^{4}p_{+}\di^{4}p_{-} \;
\delta(p_{+}^{2}-m^{2})\delta(p_{-}^{2}-m^{2})\bar{u}_{s}(p_{-})\gamma^{\mu}
u_{r}(p_{+})\widehat{a}_{s}^{\dagger}(p_{-})\widehat{a}_{r}(p_{+}) \theta(p^0_+)\theta(p^0_-)\;,
\end{align}
for the particle terms only; the other terms, i.e. those involving anti-particles and the mixed ones do not contribute 
to the connected mean values:
\begin{align}\label{connected}
\left\langle \widehat{a}_{r}^{\dagger}(p_{+})\widehat{a}_{s}(p_{-}),\widehat{T}^{\mu\nu}(0)
\right\rangle _{c,\text{GE}} & = \frac{1}{2(2\pi)^3}\bar{u}_{s}(p_{-})(\gamma^{\mu}p^{\nu}+\gamma^{\nu}p^{\mu})
u_{r}(p_{+})n_{F}(x,p_{+})(1-n_{F}(x,p_{-}))\;, \nonumber\\
\left\langle \widehat{a}_{r}^{\dagger}(p_{+})\widehat{a}_{s}(p_{-}),\widehat{j}^{\mu}(0)
\right\rangle _{c,\text{GE}} & = \frac{1}{(2\pi)^3}\bar{u}_{s}(p_{-})\gamma^{\mu}u_{r}(p_{+})n_{F}(x,p_{+})(1-n_{F}(x,p_{-}))\;.
\end{align}
where $p=(p_++p_-)/2$ and we have used
\begin{equation*}
\left\langle \widehat{a}_{r}^{\dagger}(p_{+})\widehat{a}_{s}(p_{-}),\widehat{a}_{s^{\prime}}^{\dagger}
(p_{-}^{\prime})\widehat{a}_{r^{\prime}}(p_{+}^{\prime})\right\rangle _{c,\text{GE}} = 
4\varepsilon_{+}\varepsilon_{-}\delta^{3}({\bf p}_{-}-{\bf p}_{-}^{\prime})
\delta^{3}({\bf p}_{+}-{\bf p}_{+}^{\prime}) \delta_{rr^{\prime}}\delta_{ss^{\prime}}
n_{F}(x,p_{+})(1-n_{F}(x,p_{-}))\;,
\end{equation*}
with $\varepsilon_\pm=\sqrt{m^2+{\bf p}_\pm^2}$ being the on-shell energies.
Substituting the equation \eqref{connected} into \eqref{Gfunct} and \eqref{Hfunct} and using the relation:
\begin{equation*}
(1-\e^{\beta\cdot q})n_{F}(x,p_{+})\left( 1-n_{F}(x,p_{-}) \right)=n_{F}(x,p_{+})-n_{F}(x,p_{-})\;,
\end{equation*}
one obtains: 
\begin{align}\label{Gmunu-spin-1/2}
G^{\mu\nu}(q) = & \frac{n_{F}(x,p_+)-n_{F}(x,p_-)}{4(2\pi)^3(\beta\cdot q)}
\delta\left(p^{2}+\frac{q^{2}}{4}-m^2\right) \theta(p_{+}^{0})\theta(p_-^{0}) (\slashed{p}_- + m)(\gamma^{\mu}
p^\nu+\gamma^\nu p^\mu)(\slashed{p}_+ + m)\;, \nonumber\\
H^{\mu}(q) = & -\frac{n_F(x,p_+)-n_F(x,p_-)}{2(2\pi)^3(\beta\cdot q)}
\delta\left(p^{2}+\frac{q^{2}}{4}-m^2\right) \theta(p_+^0)\theta(p_-^0) (\slashed{p}_- + m)\gamma^{\mu}(\slashed{p}_+ + m) \;.
\end{align}
In order to calculate the spin polarization, we need to evaluate the axial-vector component of the Wigner function 
${\cal A}^\mu(x,p) = \tr(\gamma^\mu \gamma^5 W^+(x,k))$. Making use of the following trace:
\begin{equation*}
\text{tr}\left[\gamma^{\mu}\gamma^{5} (\slashed{p}_{-}+m)\gamma^{\alpha}(\slashed{p}_++m)\right]
=-4\ii \epsilon^{\mu\alpha\lambda\tau} p_{\lambda}q_{\tau}\;,
\end{equation*}
and the eqs. \eqref{Gmunu-spin-1/2},\eqref{eq:wignerfinal} and \eqref{eq:wignerfinal-zeta}, one obtains:
\begin{align}\label{axialfinal}
{\cal A}_{\text{LE}}^\mu(x,p)+{\cal A}_{\text{LE}}^{\mu(\zeta)}(x,p) =& -\ii \sum_{N=0}^{\infty}\sum_{\bar{y}(x,p)}
\frac{1}{N!} D^N_y \left\{\frac{n_{F}(x,p_{+})-n_{F}(x,p_{-})}{(2\pi)^{3}(\beta\cdot q)}\delta\left(p^{2}+
\frac{q^{2}}{4}-m^{2}\right)\theta(p_{+}^{0})\theta(p_{-}^{0})\right. \nonumber\\
&\left.\times\epsilon^{\mu\rho\lambda\tau}p_{\lambda}q_{\tau}\,\frac{n_{\nu}(y)}{|p\cdot n(y)|}\left[(g_{\rho}^{\nu}
p^{\alpha}+g_{\rho}^{\alpha}p^{\nu})\Delta\beta_{\alpha}(y,x)-2g^\nu_\rho\Delta\zeta(y,x) \right]\right\} \Bigg|_{q=0,y=\bar y(x,p)}\;.
\end{align}
As the function (in fact a distribution) to be derived in $q$ is odd in that variable, only odd terms in the series 
contribute. The leading order contribution arises from the term of $N=1$. After taking a derivative with respect to
$q$, in $q=0$, this reads:
\begin{eqnarray}\label{Amu-all}
\mathcal{A}^{\mu}_{\text{LE}}(x,p)+{\cal A}_{\text{LE}}^{\mu(\zeta)}(x,p) & \simeq & \frac{2}{(2\pi)^{3}}
\epsilon^{\mu\rho\lambda\tau}p_{\lambda} n_{F}(x,p)
[1-n_{F}(x,p)] \delta(p^{2}-m^{2})\sum_{\bar{y}(x,p)}
 \Delta_{\tau\kappa}(\bar{y})\nonumber \\
 && \hspace{-1cm} \times\left\{\frac{1}{2}(g_{\rho}^{\nu}p^{\alpha}+g_{\rho}^{\alpha}p^{\nu})\left.
 \left[\partial_{y}^{\kappa}\frac{n_{\nu}(y)}{|p\cdot n(y)|}
 \Delta\beta_{\alpha}(y,x)\right]-g_{\rho}^{\nu}\left[\partial_{y}^{\kappa}\frac{n_{\nu}(y)}{|p\cdot n(y)|}
 \Delta\zeta(y,x)\right]\right\}\right|_{y=\bar{y}(x,p)}\,.
\end{eqnarray}
The derivative with respect to $y$ acts on both $\Delta\beta_{\alpha}(y,x)$, $\Delta\zeta(y,x)$, and $n_{\nu}(y)$. However
(see Appendix \ref{Appendix:B}) the curvature of $\Sigma_{\rm D}$, which relates to the derivative of $n_{\nu}(y)$, does not 
contribute to \eqref{Amu-all} due to the presence of the antisymmetric Levi-Civita tensor. Plugging \eqref{scalar-component} 
and \eqref{Amu-all} into \eqref{spin-polarization-Wigner} the final expression of the spin polarization can be derived 
(see Appendix \ref{Appendix:B} for the details) at the leading order, which reads:
\be\label{Spinfinal}
 S^{\mu}(p) \simeq -\frac{1}{8mN_p}\int_{\Sigma_{\rm D}}\!\! \di \Sigma(x) \cdot p\,
n_{F}(1-n_{F}) \epsilon^{\mu\nu\rho\lambda}p_{\nu} \sum_{\bar{y}(x,p)}\text{sgn}[p\cdot n(\bar{y})]
\bigg\{\varpi_{\rho\lambda}(\bar{y}) +\frac{2 n_{\rho}(\bar{y})}{p\cdot n(\bar{y})}\left[\xi_{\lambda\alpha}(\bar{y})
p^{\alpha}-\partial_\lambda^{\bar{y}}\zeta(\bar{y})\right]\bigg\}\;.
\ee
where $N_p=\int_{\Sigma_{\rm D}}\!\! \di\Sigma\cdot p\,\theta(p^{0})n_{F}(x,p)$ is the total particle number. 
The three terms in eq. \eqref{Spinfinal} can be identified as contributions of the thermal vorticity $\varpi^{\mu\nu}
\equiv(\partial^\nu\beta^\mu-\partial^\mu\beta^\nu)/2$, the thermal shear tensor $\xi^{\mu\nu}\equiv(\partial^\nu\beta^\mu+\partial^\mu\beta^\nu)/2$, and the spin Hall effect \citep{Liu:2020dxg} 
proportional to $\partial \zeta$, respectively, while the curvature of $\Sigma_{\rm D}$ does not contribute at this order.

The equation \eqref{Spinfinal} improves upon the formulae known in the literature in a twofold manner. First, it gives a 
new expression for the shear-induced polarization. In contrast to previous derivations, the factor $\hat{t}^{\mu}/p\cdot\hat{t}$ 
in refs. \citep{Becattini:2021iol,Becattini:2021suc} or $u^{\mu}/p\cdot u$ in refs. \citep{Fu:2021pok,Liu:2021uhn} 
(where $\hat{t}^{\mu}$ is the unit time vector in the QGP frame and $u^{\mu}$ is the fluid velocity), obtained as the result 
of significant geometric approximations, is replaced by $n^{\mu}/|p\cdot n|$. The new formula comes down to ref. \cite{Becattini:2021iol}
if $\Sigma_{\rm D}$ is a hyperplane, and to ref. \cite{Liu:2021uhn} if the velocity field is normal to $\Sigma_{\rm D}$, hence 
it reproduces previously obtained theoretical results under specific geometric assumptions. This result solves the long-standing 
issue of the ambiguity of shear-induced polarization formula, which was addressed in several papers \cite{Liu:2021nyg,Alzhrani:2022dpi}. 
Remarkably, in the eq. \eqref{Spinfinal}, the vorticity and the shear induced terms both 
receive contributions from the derivative of the $\beta$ field along $n_\mu$, but in the combination these contributions cancel 
out according to the general features of the new formula \eqref{eq:wignerfinal}. Indeed, by using eq. \eqref{ddecomp}:
\begin{align*}
 & \epsilon^{\mu\nu\rho\lambda}p_{\nu}\left[ \varpi_{\rho\lambda}+\frac{2 n_{\rho}\xi_{\lambda\alpha}
p^{\alpha}}{p\cdot n}\right] 
 = \epsilon^{\mu\nu\rho\lambda}p_{\nu} \left[ \frac{1}{2} \partial_\lambda \beta_\rho - \frac{1}{2} \partial_\rho 
 \beta_\lambda + \frac{n_{\rho} (\partial_\lambda \beta_\alpha + \partial_\alpha \beta_\lambda) p^{\alpha}}{p\cdot n} \right]
 \\
 = & \epsilon^{\mu\nu\rho\lambda}p_{\nu} \left[ - \partial_\rho \beta_\lambda + 
 \frac{n_{\rho} (\partial_\lambda \beta_\alpha + \partial_\alpha \beta_\lambda) p^{\alpha}}{p\cdot n} \right] 
 = \epsilon^{\mu\nu\rho\lambda}p_{\nu} \left[ - n_\rho D \beta_\lambda - \partial_{T\rho} \beta_\lambda + 
 \frac{n_{\rho} (n_\lambda D \beta_\alpha + n_\alpha D \beta_\lambda + \partial_{T\lambda} \beta_\alpha + \partial_{T\alpha} \beta_\lambda) p^{\alpha}}{p\cdot n} \right] \\
 = & \epsilon^{\mu\nu\rho\lambda}p_{\nu} \left[ - n_\rho D \beta_\lambda - \partial_{T\rho} \beta_\lambda + 
 n_\rho D \beta_\lambda + \frac{n_{\rho} (\partial_{T\lambda} \beta_\alpha + \partial_{T\alpha} 
 \beta_\lambda) p^{\alpha}}{p\cdot n} \right] \\
 = & \epsilon^{\mu\nu\rho\lambda}p_{\nu} \left[ - \partial_{T\rho} \beta_\lambda + 
 \frac{n_{\rho} (\partial_{T\lambda} \beta_\alpha + \partial_{T\alpha} \beta_\lambda) p^{\alpha}}{p\cdot n} \right]\;,
\end{align*}
where we have used the antisymmetric properties of the Levi-Civita symbol. The last expression apparently shows 
that the precise combination of thermal vorticity and thermal shear in equation \eqref{Spinfinal} implies
the cancellation of the derivatives along the direction $n_\rho$ perpendicular to the hypersurface. This occurs 
only for a definite value of the coefficient of thermal-shear-induced polarization in eq. \eqref{Spinfinal}; 
if the coefficient multiplying $\xi$ were different from 2, no cancellation would take place.
Particularly, when the decoupling hypersurface is isothermal and the direction of the normal vector $n_\mu$ coincides 
with $\partial_{\mu} T$, derivatives of the temperature do not contribute. Therefore, in this upgraded formalism, the 
absence of temperature gradients \cite{Becattini:2021iol,Arslan:2024dwi,Arslan:2025tan} arises naturally. Secondly, the 
thermal vorticity contribution acquires a modification by the sign function $\text{sgn}(p\cdot n)$, 
which is apparently necessary when $\Sigma_{\rm D}$ is not entirely space-like and future oriented.

The formula \eqref{Spinfinal} features additional contributions from the non-trivial solution of eq. \eqref{worldline} with 
$\bar y(x,p)\neq x$. They are geometrically related to particles moving inward across the hypersurface and traversing the fluid region \cite{Oliinychenko:2014tqa,Bugaev:1996zq,Anderlik:1998et,Grassi:2004dz}. This appears un-physical and it may be, in our 
formalism, a spurious effect generated by not having used interacting fields in the Wigner operator when dealing with in-plasma 
region in the formula \eqref{Wigner-operator}. Such additional terms need careful studies in the future but here we propose 
that they can  be discarded as follows: for the classical Cooper-Frye formula, the usual way to drop the inward-moving particles 
is to introduce a cutoff $\theta(p\cdot n)$ \cite{Bugaev:1996zq,Anderlik:1998et,Grassi:2004dz,Huovinen:2012is}. Likewise, 
introducing $\theta(p\cdot n(x))\theta(p\cdot n(\bar y))$ in eq. \eqref{Spinfinal} will remove the additional 
intersections in eq. \eqref{Spinfinal}.

It should also be stressed that the formula \eqref{Spinfinal} only contains the leading order derivatives of the normal
vector. In fact, the full formula at the leading order in the gradients of $\beta$ and $\zeta$ must include all odd-order
derivatives of the geometry-dependent factor. In this improved approximation, the \eqref{axialfinal} yields:
\begin{align}\label{axialleading}
\mathcal{A}_\text{LE}^\mu(x,p)+\mathcal{A}_\text{LE}^{\mu(\zeta)}(x,p)
\simeq&\epsilon^{\mu\rho\lambda\gamma}p_\lambda\sum_{\bar{y}(x,p)}\sum_{K=0}^\infty \frac{(-i)^{K+2}}{(K+1)!}\left[\left(g_\rho^\nu p^\alpha+g_\rho^\alpha p^\nu\right)\partial_{\tau}^y\beta_\alpha(y)-2g^\nu_\rho\partial_{\tau}^y\zeta(y)\right]\bigg|_{y=\bar{y}(x,p)}\nonumber\\
&\times\partial_{\sigma_0}^q\partial_{\sigma_1}^q\dots\partial_{\sigma_K}^q\left[\frac{n_F(x,p_+)-n_F(x,p_-)}{(2\pi)^3(\beta\cdot q)}\delta\left(p^2+\frac{q^2}{4}-m^2\right)\theta(p_+^0)\theta(p_-^0)q_\gamma\right]\bigg|_{q=0}\nonumber\\
&\times\sum_{j=0}^K  \mathcal{L}^{\sigma_0}_y\dots
\mathcal{L}^{\sigma_{j-1}}_y\Delta^{\sigma_j \tau}(y)\mathcal{L}^{\sigma_{j+1}}_y\dots\mathcal{L}^{\sigma_K}_y\frac{n_\nu(y)}{|p\cdot n(y)|}\bigg|_{y=\bar{y}(x,p)}\,,
\end{align}
where $\mathcal{L}_y^\sigma\equiv \Delta^{\sigma\tau}(y)\partial_\tau^y$.
The leading correction to the formula \eqref{Spinfinal} arises from the $K=2$ term in eq. \eqref{axialleading}. At this 
order, the spin polarization receives contributions from $(\partial_\mu \partial_\nu n_\alpha)(\partial_\lambda \beta_\kappa)$, $(\partial_\mu n_\alpha)(\partial_\nu n_\rho)(\partial_\lambda \beta_\kappa)$, $(\partial_\mu \partial_\nu n_\alpha)(\partial_\lambda \zeta)$, and $(\partial_\mu n_\alpha)(\partial_\nu n_\rho)(\partial_\lambda \zeta)$. Explicit expressions 
for these effects are lengthy and they are not shown in this paper. Their impacts on global and local spin polarization 
in heavy-ion collisions will be studied in future works.

%***************************************************************************
\section{Conclusions}
%***************************************************************************

In conclusion, we have developed a new method to calculate the Wigner function at LTE on the 
decoupling hypersurface in heavy-ion collisions. By inverting the momentum and the hypersurface 
integrals, we have been able to remove specific assumptions on the geometry of the hypersurface and
to include in the gradient expansion those of the normal vector to the hypersurface.
The space-.time derivatives in the normal direction of the hypersurface are naturally excluded, 
which has a remarkable implication in the vanishing contributions from the temperature gradient if
the hypersurface is isothermal. We have derived upgraded expressions of the spin polarization
of spin 1/2 fermions, which merits verifications in numerical simulations.

\begin{acknowledgments}
The authors thank G.-L. Ma and Q. Wang for insightful discussions. This work is supported in part by the 
Italian Ministry of University and Research, project PRIN2022 “Advanced probes of the Quark Gluon Plasma”, 
Next Generation EU, Mission 4 Component 1. The data that support the findings of this article are openly 
available \cite{sheng_2026_18336680}.
\end{acknowledgments}

\bibliographystyle{apsrev}
\bibliography{main}

\begin{appendix}
%************************************************************************
\section{Compact form of Wigner function} \label{Appendix:A}
%************************************************************************

Plugging the eq. \eqref{eq:I-n-final} into the eq. \eqref{eq:W-LTE-In}, 
we obtain:
\begin{eqnarray}
\Delta W_{\text{LE}}^{+}(x,p) & = & \frac{1}{|p^{0}|}\sum_{N=0}^{\infty}\frac{(-i)^{N}}{N!}\left.
\left[\partial^{q}_{\nu_{1}}\partial^{q}_{\nu_{2}}\cdots\partial^{q}_{\nu_{N}}
G^{\mu\nu}(q)\right]\right|_{q=0}\nonumber \\
 &  & \times\int_{\Sigma_{\rm D}}\!\! \di\Sigma_{\mu}(y)\, \Delta\beta_{\nu}(y,x)
 \left[ \partial^{\nu_{1}}_{x}\partial^{\nu_{2}}_{x}\cdots\partial^{\nu_{N}}_{x}\delta^{3}
 \left({\bf y}-{\bf x}-\frac{\bf p}{p^{0}}(y^{0}-x^{0})\right)\right] \;.\label{eq:delta-W-1}
\end{eqnarray}
The presence of the partial derivatives of the $\delta$-function makes the integral difficult to 
evaluate. It is convenient to convert the above equation to the form in \eqref{eq:delta-W-2} through
the following steps. First, we split the partial derivative into the sum of two different operators
$$
 \partial_x = \partial^{(1)}_{x} + \partial^{(2)}_{x}\;,
$$
acting only on the $\delta$-function and $\Delta\beta_{\nu}(y,x)$, respectively. Thereby, the 
partial derivatives in eq. \eqref{eq:delta-W-1} can be recast as 
$\left[\partial^{q}\cdot\partial^{(1)}_{x}\right]^{N}$
and taken out of the integral. The latter can be expressed as follows, by means of the Binomial theorem:
$$
\left[\partial^{q}\cdot\partial^{(1)}_x\right]^{N} = \left[\partial^{q}\cdot(\partial^{(1)}_x+
\partial^{(2)}_x)-\partial^{q}\cdot\partial_{(2)}^{x}\right]^{N}
  = \sum_{M=0}^{N}\frac{N!(-1)^{M}}{M!(N-M)!}\left[\partial^{q}\cdot \partial_{x} \right]^{N-M}
  \left[\partial^{q}\cdot\partial^{(2)}_{x}\right]^{M}\;,
$$
The derivative $\partial_x^{(2)}$ only acts on the $\Delta\beta$ function and can be moved into the
integral, yielding the expression in the eq. \eqref{eq:delta-W-2}.

In the eq. \eqref{eq:delta-W-2}, the space-time point $y$ is located on $\Sigma_{\rm D}$. As 
described in the main text, $\Sigma_{\rm D}$ can be split into several branches, where in each 
the decoupling time $y^0$ is a single-valued function of ${\bf y}$, i.e., $y^{0}=f_{k}({\bf y})$. 
Then the equation \eqref{deltas} follows because, on each branch:
\begin{equation}
\delta^{3}\left({\bf y}-{\bf x}- \frac{\bf p}{p^0}(y^{0}-x^{0}) \right)=
\sum_{k,i}\frac{|p^{0}|}{|p\cdot\sigma(\bar{y}_{k,i})|}\delta^{3}({\bf y}-\bar{{\bf y}}_{k,i})\;,
\end{equation}
where $\bar{y}_{k,i}^{\mu}(x)$ denotes the $i$-th crossing point of the worldline \eqref{worldline}
with the $k$-th branch of $\Sigma_{\rm D}$. The pre-factor on the right hand side is the 
inverse of the determinant of the matrix:
$$
\frac{\partial}{\partial y^j}\left[{\bf y}-{\bf x}- \frac{\bf p}{p^0}(f_k({\bf y})-x^{0})\right]^i =
\delta^i_j - \frac{p^i}{p^0} \frac{\partial f_k({\bf y})}{\partial y^j} = \delta^i_j - \frac{p^i}{p^0} 
\sigma^j = \frac{p^0 \delta^i_j - p^i \sigma^j}{p^0} \;.
$$
Carrying out the integration in $\di^3 \y$, and using the relations \eqref{deltas}, the equation 
\eqref{eq:delta-W-2} is converted to eq. \eqref{eq:delta-W-3}.

In eq. \eqref{chain-rule}, the derivative of $\bar y(x,p)$ with respect to $x$ is obtained by taking into account
that $\bar{y}$ locates on the world line:
\begin{equation}
\bar{{\bf y}}={\bf x}+\frac{{\bf p}}{p^{0}}(\bar{y}^{0}-x^{0})\;.
\end{equation}
Taking partial derivatives with respect to $x^\mu$
of the above equation:
\begin{equation}
\frac{\partial\bar{y}^{j}}{\partial x^{\mu}}-\frac{p^{j}}{p^{0}}\frac{\partial\bar{y}^{0}}
{\partial\bar{y}^{l}}\frac{\partial\bar{y}^{l}}{\partial x^{\mu}} = 
\left[ \delta^j_l + \frac{p^j}{p^0} \sigma_l(\bar{y}) \right] \frac{\partial\bar{y}^{l}}{\partial x^{\mu}} 
= \delta_{\mu}^{j}-\frac{p^{j}}{p^{0}}\delta_{\mu}^{0}\;,
\end{equation}
where $j,l=1,2,3$ and where $\sigma_{\mu}(\bar{y})$ is the normal vector of $\Sigma_{\rm D}$ at 
the spacetime point $\bar{y}$. For a point $\bar{y}=\bar{y}_{k,i}$, which denotes the $i$-th solution of $\bar{{\bf y}}={\bf x}+\frac{{\bf p}}{p^{0}}(\bar{y}^{0}-x^{0})$ on the $k$-th branch of the hypersurface, we obtain:
\begin{equation}
\sigma_{\mu}(\bar{y}_{k,i})=\left(1,-\frac{\partial f_{k}(\bar{{\bf y}}_{k,i})}
{\partial\bar{{\bf y}}_{k,i}}\right)\;.
\end{equation}
The $3 \times 3$ matrix:
$$
  A^j_l = \left[ \delta^j_l + \frac{p^j}{p^0} \sigma_l(\bar{y}_{k,i}) \right]
$$
can be inverted and one obtains:
$$
\frac{\partial\bar{y}_{k,i}^{j}}{\partial x^{\mu}} = A^{-1j}_{\;\;\; l} \left(\delta_{\mu}^{l}-
\frac{p^{l}}{p^{0}}\delta_{\mu}^{0}\right) = \left[ \delta^j_l - \frac{p^j \sigma_l(\bar{y}_{k,i}) }{p \cdot \sigma(\bar{y}_{k,i}) } \right]  
\left(\delta_{\mu}^{l}-\frac{p^{l}}{p^{0}}g_{\mu}^{0}\right) \;.
$$
If $\mu = 0$ we then have:
$$
\frac{\partial\bar{y}_{k,i}^{j}}{\partial x^{0}} = -\frac{p^j}{p \cdot \sigma(\bar{y}_{k,i}) }\;,
$$
while if $\mu = m$ spacial index
$$
\frac{\partial\bar{y}_{k,i}^{j}}{\partial x^{m}} = \delta^j_m - \frac{p^j \sigma_m(\bar{y}_{k,i}) }{p \cdot \sigma(\bar{y}_{k,i}) } \;,
$$
so that, altogether:
\be\label{deriv1}
 \frac{\partial\bar{y}_{k,i}^{j}}{\partial x^{\mu}} = \delta^j_\mu - 
 \frac{p^j \sigma_\mu(\bar{y}_{k,i}) }{p \cdot \sigma(\bar{y}_{k,i}) }\;.
\ee
Similarly, one can calculate the derivative of $y^0$ with respect to $x$:
$$
\frac{\partial\bar{y}_{k,i}^{0}}{\partial x^{\mu}} = \frac{\partial\bar{y}_{k,i}^{0}}
{\partial\bar{y}_{k,i}^{j}}\frac{\partial\bar{y}_{k,i}^{j}}{\partial x^{\mu}} = 
-\sigma_j(\bar{y}_{k,i})  \left[ \delta^j_\mu - \frac{p^j \sigma_\mu(\bar{y}_{k,i}) }{p \cdot \sigma(\bar{y}_{k,i}) } \right]\;,
$$
whence we obtain:
\be\label{deriv2}
 \frac{\partial\bar{y}_{k,i}^{0}}{\partial x^{0}} = \sigma_0(\bar{y}_{k,i})  - \frac{p^0}{p \cdot \sigma(\bar{y}_{k,i}) } \qquad\qquad
 \frac{\partial\bar{y}_{k,i}^{0}}{\partial x^{m}} = - \frac{p^0 \sigma_m(\bar{y}_{k,i}) }{ p \cdot \sigma(\bar{y}_{k,i}) } \;.
\ee
The equations \eqref{deriv1} and \eqref{deriv2} can be written in a compact form as:
\begin{equation}
 \frac{\partial\bar{y}_{k,i}^{\nu}}{\partial x^{\mu}}=\delta_{\mu}^{\nu}-\frac{p^{\nu}
 \sigma_{\mu}(\bar{y}_{k,i})}{p\cdot\sigma(\bar{y}_{k,i})} \equiv \Delta_{\mu}^{\;\nu}(\bar{y}_{k,i})\;,
\end{equation}
so that eq. \eqref{chain-rule} can be rewritten as:
\begin{equation}
\frac{\di}{\di x^\mu} g(x,\bar{y}_{k,i}(x))= \frac{\partial}{\partial x^\mu} g(x,\bar{y}_{k,i}(x))+ 
\Delta^{\ \nu}_\mu(\bar{y}_{k,i}) \frac{\partial g(x,y)}{\partial y^\nu}\Bigg|_{y=\bar{y}_{k,i}(x,p)}\;.
\end{equation}
By using the last result, the eq. \eqref{eq:delta-W-3} can be rewritten as:
\begin{eqnarray}
\Delta W_{\text{LE}}^{+}(x,p) & = &\sum_{N=0}^{\infty} \sum_{k,i} s_k \frac{(-i)^{N}}{N!}
\left.\left[\partial^{q}_{\nu_{1}}\cdots\partial^{q}_{\nu_{N}}G^{\mu\nu}(q)\right]\right|_{q=0} 
\sum_{M=0}^{N}\frac{N!(-1)^{M}}{M!(N-M)!}\left[\partial^{\nu_{M+1}}_{x}+
\Delta^{\nu_{M+1}\rho_{M+1}}(\bar{y}_{k,i})\partial_{\rho_{M+1}}^{y}\right]\nonumber \\
 &  & \times\left.\cdots\left[\partial^{\nu_{N}}_{x}+\Delta^{\nu_N\rho_{N}}(\bar{y}_{k,i})
 \partial_{\rho_{N}}^{y}\right]\partial^{\nu_{1}}_{x}\cdots\partial^{\nu_{M}}_{x}
 \frac{\sigma_{\mu}(y)}{|p\cdot\sigma(y)|}\Delta\beta_{\nu}(y,x)\right|_{y=\bar{y}_{k,i}}\;.
\end{eqnarray}
Using again the Binomial theorem, the last expression can be recast as:
\begin{equation}
\Delta W_{\text{LE}}^{+}(x,p)=\sum_{N=0}^{\infty}\sum_{k,i} s_k \frac{(-i)^{N}}{N!}
\left.\left[\Delta^{\alpha\rho}(\bar{y}_{k,i})\partial_{\rho}^{y}\partial_{\alpha}^{q}
\right]^{N} G^{\mu\nu}(p,q)\frac{\sigma_{\mu}(y)}{|p\cdot\sigma(y)|}\Delta\beta_{\nu}(y,x)
\right|_{q=0,y=\bar{y}_{k,i}}\;.
\end{equation}
Now, because of its definition (see text below eq. \eqref{eq:normal-vector}), the sign $s_k$ 
is such that:
$$
  s_k \sigma_\mu = n_\mu \sqrt{|\sigma \cdot \sigma|}\;,
$$
where $n_\mu$ is the outward pointing unit vector normal to the hypersurface $\Sigma_{\rm D}$. 
Besides:
$$
\frac{n_\mu \sqrt{|\sigma \cdot \sigma|}}{|p \cdot \sigma|} = \frac{n_\mu}{|p \cdot n|}\;,
$$
and
$$
 \Delta^{\ \nu}_\mu = \delta_{\mu}^{\nu}-\frac{p^{\nu}
 \sigma_{\mu}}{p\cdot\sigma} = \delta_{\mu}^{\nu}-\frac{p^{\nu}n_{\mu}}{p\cdot n}\;.
$$
By taking the above equalities into account and introducing the operator $D_y$ in eq. 
\eqref{eq:operator-D-q-y}, we can express  $\Delta W_{\rm LE}^+$ as shown in eq. \eqref{eq:wignerfinal}. 
Note the argument of the operator $\Delta^{\alpha\rho}$ is $\bar{y}$, such that the 
derivative with respect to $y$ does not act on $\Delta^{\alpha\rho}$.

%******************************************************************************
\section{Spin polarization for spin-1/2 fermions}\label{Appendix:B}
%******************************************************************************
In eq. \eqref{Amu-all}, the contribution from $\Delta\beta$ reads:
\begin{eqnarray}\label{Amu-Delta-beta}
\mathcal{A}^{\mu}_{\text{LE}}(x,p) & \simeq & \frac{2}{(2\pi)^{3}}n_{F}(x,p)[1-n_{F}(x,p)]
\delta(p^{2}-m^{2})\epsilon^{\mu\rho\lambda\tau}p_{\lambda}\nonumber \\
 &  & \times\frac{1}{2}(g_{\rho}^{\nu}p^{\alpha}+g_{\rho}^{\alpha}p^{\nu})\sum_{\bar{y}(x,p)}
 \Delta_{\tau\kappa}(\bar{y})\left.\left[\partial_{y}^{\kappa}\frac{n_{\nu}(y)}{|p\cdot n(y)|}
 \Delta\beta_{\alpha}(y,x)\right]\right|_{y=\bar{y}(x,p)}\;.
\end{eqnarray}
The derivative with respect to $y$ acts on both $\Delta\beta_{\alpha}(y,x)$ and $ n_{\nu}(y)$,
giving rise to two terms. For the former, one obtains a factor which is proportional to:
\begin{eqnarray*}
&& \frac{1}{2}\epsilon^{\mu\rho\lambda\tau}p_{\lambda}\Delta_{\tau\kappa}(\bar{y}) n_{\nu}(\bar{y})(g_{\rho}^{\nu}p^{\alpha}+g_{\rho}^{\alpha}p^{\nu})\left[\left.\partial_{y}^{\kappa}
\beta_{\alpha}(y)\right|_{y=\bar{y}(x,p)}\right]\nonumber \\
& = & \frac{1}{2}\epsilon^{\mu\nu\lambda\tau}p_{\lambda}\Delta_{\tau\kappa}(\bar{y})
 n_{\nu}(\bar{y})p^{\alpha}\left[\left.\partial_{y}^{\kappa}\beta_{\alpha}(y)
\right|_{y=\bar{y}(x,p)}\right]+\left[p\cdot n(\bar{y})\right]\frac{1}{2}\epsilon^{\mu\alpha\lambda\tau}
p_{\lambda}\Delta_{\tau\kappa}(\bar{y})\left[\left.\partial_{y}^{\kappa}\beta_{\alpha}(y)
\right|_{y=\bar{y}(x,p)}\right]\;.
\end{eqnarray*}
Plugging the explicit form of $\Delta_{\tau\kappa}(\bar{y})$ in eq. \eqref{operatorDelta} it 
becomes:
\begin{equation*}
\frac{1}{2}\epsilon^{\mu\nu\lambda\kappa}p_{\lambda} n_{\nu}(\bar{y})p^{\alpha}
\left[\left.\partial_{\kappa}^{y}\beta_{\alpha}(y)\right|_{y=\bar{y}(x,p)}\right] 
 +[p\cdot n(\bar{y})]\frac{1}{2}\epsilon^{\mu\alpha\lambda\kappa}p_{\lambda}
 \left[\left.\partial_{\kappa}^{y}\beta_{\alpha}(y)\right|_{y=\bar{y}(x,p)}\right]-
 \frac{1}{2}\epsilon^{\mu\alpha\lambda\tau}p_{\lambda}p^{\kappa} n_{\tau}(\bar{y})
 \left[\left.\partial_{\kappa}^{y}\beta_{\alpha}(y)\right|_{y=\bar{y}(x,p)}\right]\;,
\end{equation*}
whence, renaming repeated indices:
\be\label{Amu-LE-1}
 [p\cdot n(\bar{y})]\frac{1}{2}\epsilon^{\mu\nu\rho\lambda}p_{\nu}\left[\left.\partial_{\rho}^{y}
 \beta_{\lambda}(y)\right|_{y=\bar{y}(x,p)}\right] -\frac{1}{2}\epsilon^{\mu\nu\rho\lambda}
 p_{\nu} n_{\rho}(\bar{y})p^{\alpha}\left[\left.\partial_{\lambda}^{y}\beta_{\alpha}(y)+
 \partial_{\alpha}^{y}\beta_{\lambda}(y)\right|_{y=\bar{y}(x,p)}\right]\;.
\ee
By using the definitions of thermal vorticity and thermal shear:
\begin{eqnarray*}
\varpi_{\rho\lambda} & = & -\frac{1}{2}\left(\partial_{\rho} \beta_{\lambda}-
\partial_{\lambda}\beta_{\rho} \right)\;, \nonumber \\
\xi_{\rho\lambda} & = & \frac{1}{2}\left(\partial_{\rho} \beta_{\lambda} +
\partial_{\lambda}\beta_{\rho} \right)\;,
\end{eqnarray*}
in the eq. \eqref{Amu-LE-1} and reinstating the missing factors, we obtain the first part 
of $\mathcal{A}_{\text{LE}}^{\mu}$:
\be\label{eq:axial-vector}
\mathcal{A}_{\text{LE}-1}^\mu (x,p) = 
-\frac{\delta(p^{2}-m^{2})}{(2\pi)^{3}}n_{F}(x,p)(1-n_{F}(x,p))\sum_{\bar{y}(x,p)}
\text{sgn}[p\cdot n(\bar{y})]
 \epsilon^{\mu\nu\rho\lambda}p_{\nu}\left[\varpi_{\rho\lambda}(\bar{y})+
 \frac{2}{p\cdot n(\bar{y})} n_{\rho}(\bar{y})\xi_{\lambda\alpha}(\bar{y})p^{\alpha}\right]\;.
\ee
On the other hand, the contribution from the derivative of the normal vector $ n$
is proportional to
\begin{equation}\label{eq:Amu-LE-2}
\epsilon^{\mu\rho\lambda\tau}p_{\lambda}(g_{\rho}^{\nu}p^{\alpha}+g_{\rho}^{\alpha}p^{\nu})
\Delta_{\tau\kappa}(\bar{y})\Delta\beta_{\alpha}(\bar{y},x)\left.\left[\partial_{y}^{\kappa}
\frac{ n_{\nu}(y)}{|p\cdot n(y)|}\right]\right|_{y=\bar{y}}\;,
\end{equation}
which in turn gives rise to two terms. The first, obtained by contracting with $p^\nu$, is 
proportional to: 
\begin{equation*}
p^{\nu}\left.\left[\partial_{y}^{\kappa}\frac{ n_{\nu}(y)}{|p\cdot n(y)|}\right]
\right|_{y=\bar{y}}=\left.\left[\partial_{y}^{\kappa}\text{sgn}[p\cdot n(y)]\right]
\right|_{y=\bar{y}}\propto\delta( p\cdot n(\bar y))\;,
\end{equation*}
which vanishes unless $p\cdot n(\bar{y})=0$. In our derivation, in order to ensure the
that the $\delta$ function can be solved in terms of the ${\bf y}$ variable, we had to assume
that $p\cdot n(\bar{y})=0$ so we neglect the above term. Therefore, we can rewrite the
above term, without changing it, as:
\begin{equation}\label{eq:Amu-LE-2-step-1}
 \epsilon^{\mu\rho\lambda\tau}p_{\lambda}
(g_{\rho}^{\nu}p^{\alpha}-g_{\rho}^{\alpha}p^{\nu})\Delta_{\tau\kappa}(\bar{y})
\Delta\beta_{\alpha}(\bar{y},x)\left.\left[\partial_{y}^{\kappa}
\frac{ n_{\nu}(y)}{|p\cdot n(y)|}\right]\right|_{y=\bar{y}}\;,
\end{equation}
that is by changing the sign of the second term in the bracket. Using the Schouten identity:
\begin{equation}\label{Schouten-identity}
\epsilon^{\mu\nu\lambda\tau}p^{\alpha}+\epsilon^{\nu\lambda\tau\alpha}p^{\mu}+
\epsilon^{\lambda\tau\alpha\mu}p^{\nu}+\epsilon^{\tau\alpha\mu\nu}p^{\lambda}+
\epsilon^{\alpha\mu\nu\lambda}p^{\tau}=0\;,
\end{equation}
we have:
$$
\epsilon^{\mu\rho\lambda\tau}p_{\lambda}(g_{\rho}^{\nu}p^{\alpha}-g_{\rho}^{\alpha}p^{\nu})  =  (\epsilon^{\mu\nu\lambda\tau}p^{\alpha}+\epsilon^{\lambda\tau\alpha\mu}p^{\nu})p_{\lambda}
 = -(\epsilon^{\nu\lambda\tau\alpha}p^{\mu}+\epsilon^{\tau\alpha\mu\nu}p^{\lambda}
 +\epsilon^{\alpha\mu\nu\lambda}p^{\tau})p_{\lambda}\;,
$$
Plugging this relation into eq. \eqref{eq:Amu-LE-2-step-1} and taking into account that  
$p^{\tau}\Delta_{\tau\kappa}(\bar{y})=0$, the expression in \eqref{eq:Amu-LE-2-step-1} can be 
rewritten as:
\begin{eqnarray}\label{eq:Amu-LE-2-step-2}
&& -(\epsilon^{\nu\lambda\tau\alpha}p^{\mu}+\epsilon^{\tau\alpha\mu\nu}p^{\lambda})p_{\lambda}
\Delta_{\tau\kappa}(\bar{y})\Delta\beta_{\alpha}(\bar{y},x)\left.\left[\partial_{y}^{\kappa}
\frac{ n_{\nu}(y)}{|p\cdot n(y)|}\right]\right|_{y=\bar{y}}\nonumber \\
 & = & -\epsilon^{\nu\lambda\tau\alpha}(p^{\mu}p_{\lambda}-g_{\lambda}^{\mu}p^{2})
 \Delta_{\tau\kappa}(\bar{y})\Delta\beta_{\alpha}(\bar{y},x)\left.\left[\partial_{y}^{\kappa}
 \frac{ n_{\nu}(y)}{|p\cdot n(y)|}\right]\right|_{y=\bar{y}}\;.
\end{eqnarray}
With the help of the definition of $\Delta_{\tau\kappa}(\bar{y})$, one can prove that:
\begin{eqnarray}\label{eq:schouten-2}
\epsilon^{\nu\lambda\tau\alpha}\Delta_{\tau}^{\ \kappa}(\bar{y}) & = & 
\epsilon^{\nu\lambda\tau\alpha}\left[g_{\tau}^{\kappa}-\frac{p^{\kappa} n_{\tau}(\bar{y})}
{p\cdot n(\bar{y})}\right] =  \left(p^{\tau}\epsilon^{\nu\lambda\kappa\alpha}-p^{\kappa}\epsilon^{\nu\lambda\tau\alpha}\right)\frac{ n_{\tau}(\bar{y})}
{p\cdot n(\bar{y})}\nonumber \\
 & = & -\left(p^{\nu}\epsilon^{\lambda\kappa\alpha\tau}+p^{\lambda}\epsilon^{\kappa\alpha\tau\nu}
 +p^{\alpha}\epsilon^{\tau\nu\lambda\kappa}\right)\frac{ n_{\tau}(\bar{y})}
 {p\cdot n(\bar{y})}\;,
\end{eqnarray}
where we have again used the Schouten identity \eqref{Schouten-identity} in the last step. 
Substituting \eqref{eq:schouten-2} into eq. \eqref{eq:Amu-LE-2-step-2} and noting that
(see also above discussion):
\begin{eqnarray*}
&&p^{\lambda}(p^{\mu}p_{\lambda}-g_{\lambda}^{\mu}p^{2})=0\;, \\
&& p^{\nu}\left.\left[\partial_{y}^{\kappa}\frac{ n_{\nu}(y)}{|p\cdot n(y)|}\right]
\right|_{y=\bar{y}}=\left.\left[\partial_{y}^{\kappa}\text{sgn}[p\cdot n(y)]\right]
\right|_{y=\bar{y}} = 0\;,
\end{eqnarray*}
the expression in eq. \eqref{eq:Amu-LE-2-step-2} gets its final form:
\begin{equation}\label{finalform}
(p^{\mu}p_{\lambda}-g_{\lambda}^{\mu}p^{2})\epsilon^{\lambda\tau\nu\kappa}
\frac{ n_{\tau}(\bar{y})}{p\cdot n(\bar{y})}\left.\left[\partial_{\kappa}^{y}
\frac{ n_{\nu}(y)}{|p\cdot n(y)|}\right]\right|_{y=\bar{y}}
p^{\alpha}\Delta\beta_{\alpha}(\bar{y},x)\;.
\end{equation}
Defining the new vector:
\begin{equation}\label{defOmega}
\Omega^{\lambda}(\bar{y})\equiv\epsilon^{\lambda\tau\nu\kappa}\frac{ n_{\tau}(\bar{y})}{p\cdot n(\bar{y})}\left.\left[\partial_{\kappa}^{y}\frac{ n_{\nu}(y)}{|p\cdot n(y)|}\right]\right|_{y=\bar{y}}=\frac{\text{sgn}[p\cdot n(\bar{y})]}{[p\cdot n(\bar{y})]^{2}}\epsilon^{\lambda\tau\nu\kappa} n_{\tau}(\bar{y})\left.
\left[\partial_{\kappa}^{y} n_{\nu}(y)\right]\right|_{y=\bar{y}}\;,
\end{equation}
which depends on the curvature of the decoupling hypersurface at point $\bar{y}$, the
equation \eqref{finalform} can be rewritten as:
\begin{equation}
 (p^{\mu}p_{\lambda}-g_{\lambda}^{\mu}p^{2}) 
 p^{\alpha}\Delta\beta_{\alpha}(\bar{y},x) \Omega^{\lambda}(\bar{y})\;.
\end{equation}
However, according to the definition of the normal unit vector we have:
$$
   n_\mu = A \sigma_\mu\;,
$$
with $A$ suitable scalar and $\sigma$ like in eq. \eqref{eq:normal-vector}. Hence:
\be\label{dersigma}
 \partial_{\kappa} n_{\nu} = A \, \partial_\kappa \sigma_\nu + (\partial_\kappa A) n_\nu\;.
\ee
Since:
\begin{equation*}
\partial_{\kappa}^{y} \sigma_{\nu}(y)= \begin{cases}
0 & \kappa=0\text{ or }\nu=0\;,\\
-\frac{\partial f_{k}({\bf y})}{\partial y^{i}\partial y^{j}} & \kappa=i,\ \nu=j,\ i,j=1,2,3\;,
\end{cases}
\end{equation*}
it turns out that the gradient of $\sigma$ in eq. \eqref{dersigma} is symmetric under the 
exchange of its indices. Therefore, the contraction with $\epsilon^{\lambda\tau\nu\kappa}$ in 
equation \eqref{defOmega} of both the terms in equation \eqref{dersigma} vanishes and so $\Omega^{\lambda}=0$. In conclusion, the expression in \eqref{finalform} vanishes, demonstrating that the curvature of $\Sigma_{\rm D}$ does not contribute to the spin polarization at the 
leading order. 

On the other hand, the contribution from $\Delta\zeta$ is given by:
\begin{eqnarray}\label{eq:Amu-zeta}
{\cal A}_{\text{LE},\zeta}^\mu(x,p) &=& -\frac{2}{(2\pi)^3}\sum_{\bar{y}(x,p)}n_F(x,p)[1-n_F(x,p)]\delta(p^2-m^2) \nonumber\\
&&\times\epsilon^{\mu\nu\lambda\tau}p_{\lambda}\,
\Delta_{\tau\kappa}(\bar{y})\left[\partial _y^\kappa\frac{n_{\nu}(y)}{|p\cdot n(y)|} \Delta\zeta(y,x) \right] \Bigg|_{q=0,y=\bar y(x,p)}\,.
\end{eqnarray}
Using eq. \eqref{eq:schouten-2}, one can prove that:
\begin{eqnarray}
\epsilon^{\mu\nu\lambda\tau}p_\lambda\Delta_{\tau\kappa}\left.\left[\partial_y^\kappa\frac{n_\nu(y)}{|p\cdot n(y)|}\right]\right|_{y=\bar{y}}&=&(p^\mu p_\lambda-p^2 g^\mu_\lambda)\epsilon^{\lambda\nu\kappa \tau}\frac{n_\tau(\bar{y})}{p\cdot n(\bar{y})}\left.\left[\partial_\kappa^y\frac{n_\nu(y)}{|p\cdot n(y)|}\right]\right|_{y=\bar{y}} \nonumber\\
&&-\epsilon^{\lambda\kappa\tau\mu}p_\lambda\frac{n_\tau(\bar{y})}{p\cdot n(\bar{y})}p^\nu\left.\left[\partial_\kappa^y\frac{n_\nu(y)}{|p\cdot n(y)|}\right]\right|_{y=\bar{y}}\;.
\end{eqnarray}
As shown in the previous section, the terms in the above equation vanish, indicating that the curvature of $\Sigma_{\rm D}$ does not contribute to ${\cal A}_{\text{LE},\zeta}^\mu$ in eq. \eqref{eq:Amu-zeta}. On the other hand, the contribution from $\partial_y\zeta$ reads
\begin{eqnarray}\label{eq:Amu-LE-zeta-final}
{\cal A}_{\text{LE},\zeta}^\mu(x,p) &=& \frac{2}{(2\pi)^3}\sum_{\bar{y}(x,p)}n_F(x,p)[1-n_F(x,p)]\delta(p^2-m^2) \nonumber\\
&&\times\epsilon^{\mu\nu\rho\lambda}p_{\nu}\,\frac{n_\rho(\bar{y})}{|p\cdot n(\bar{y})|}\partial_\lambda^{\bar{y}}\zeta(\bar{y})\;.
\end{eqnarray}
Finally, plugging eqs. \eqref{scalar-component}, \eqref{eq:axial-vector}, and \eqref{eq:Amu-LE-zeta-final} into eq. 
\eqref{spin-polarization-Wigner}, the vorticity- and shear-induced polarization as well as the spin Hall effect in \eqref{Spinfinal} are obtained.

\end{appendix}
\end{document}